\documentclass[aps, pra, twocolumn, amssymb, amsmath, amsfonts, showpacs, superscriptaddress, preprintnumbers, a4paper]{revtex4-2} 
\usepackage{amssymb, amsmath, amsthm}
\usepackage{xcolor}
\usepackage{graphicx} % Required for inserting images
\usepackage{hyperref}
\usepackage{comment}
\usepackage{physics}
\usepackage{footnote}
\usepackage[utf8]{inputenc}
\usepackage{dsfont}
\usepackage{comment}
\newcommand{\isEquivTo}[1]{\underset{#1}{\sim}}

\begin{document}

\title{Open-system quantum many-body scars: a theory} 
\author{Lorenzo Gotta}
\affiliation{Department of Quantum Matter Physics, University of Geneva, 1211 Geneva, Switzerland.}
%\date{March 2024}

\begin{abstract}
In this work, we undertake the problem of formally introducing a notion of quantum many-body scarring in open quantum systems governed by the Lindblad equation. To this goal, we rely on the commutant-algebra framework for the description of strong symmetries to introduce the unconventional strong-symmetry structure leading to the existence of anomalous stationary states, which we dub open-system quantum many-body scars (OSQMBS), besides a typical infinite-temperature state. We provide several benchmarks of the theoretical predictions on the stationary-state manifold and on convergence to stationarity, as well as describe the time-evolution of off-diagonal coherences among the Hilbert space symmetry sectors identified by OSQMBS and their orthogonal complement. Moreover, we investigate the existence of asymptotic OSQMBS (AOSQMBS), the latter being states that, despite converging to the typical infinite-temperature state in the large-time limit, display anomalously-large relaxation time scales, which we thoroughly describe by means of the behavior of their fidelity as a function of time through proper scaling Ansaetze.  
\end{abstract}

\maketitle

\section{Introduction}

Symmetries represent a long-standing cornerstone in the understanding of the properties of quantum many-body systems. While traditionally the impact of symmetries on ground-state and low-energy properties have been the object of several investigations, the recent rising interest towards the dynamical evolution of quantum many-body systems has attracted attention towards the interplay of symmetries and quantum many-body dynamics~\cite{Ogunnaike_2023, Moudgalya_2024}.

Despite the foundational importance of time evolution and equilibration in closed quantum systems~\cite{Rigol_2008}, all experimentally feasible setups involve loss of coherence in time due to the interaction between the system of interest and its surrounding environment~\cite{Daley_2014}. Thus, it is highly desirable to understand the dynamical properties of quantum many-body systems in an open setting~\cite{Fazio_2025}, by considering as a starter the weak-coupling and Markovian limit in which the time evolution is governed by Lindblad equation.

The characterization of symmetries in Lindblad open quantum systems, as well as their interplay with the stationary-state manifold of the Lindblad evolution, is largely based on the notions of weak and strong symmetries~\cite{Buca_2012,Albert_2014,Zhang_2020,Nigro_2019,Yoshida_2024}; the latter, who will be the focus of the present work, can be naively described as being linked to the existence of unitary operators which commute both with the hamiltonian and the local dissipators chosen as jump operators in the Lindblad superoperator. 

A recent algebraic formalism for treating on an equal footing both standard extensive symmetries and non-extensive symmetries that are responsible for ergodicity-breaking in closed quantum systems has been introduced and described in terms of the notion of commutant algebra of a corresponding von Neumann algebra generated by a set of local Hamiltonian terms~\cite{Moudgalya_2022,Moudgalya_2023,Moudgalya_2024_scars}. Moreover, beyond the study of closed quantum systems, a few studies have successfully applied the commutant-algebra framework for symmetries to the setting of Lindblad systems~\cite{Li_2023,Li_2025,Paszko_2025}, where the aforesaid formalism has allowed for the complete characterization of strong symmetries of families of open quantum systems in a wide variety of settings and the ensuing description of the stationary properties of such systems. 

In the present work, we further extend the application of the commutant-algebra-based formalism to Lindblad dynamics. Our main contribution consists in the formulation of a systematic theory of OSQMBS. We describe the latter as exceptional stationary states that are not protected by any standard discrete or continuous symmetry, existing besides typical infinite-temperature ones that attract all the initial states but a measure zero set of them, as first exemplified by the open quantum spin chain characterized in Ref.~\cite{Marche_2024}. We identify the algebraic conditions that open quantum systems must satisfy to host OSQMBS, which consist in certain structure properties of the commutant algebra associated to the given bond algebra, the latter defining a class of open quantum systems. Moreover, we explore the special features of systems with OSQMBS by presenting the manifold of their stationary states, diverse informations on their dynamical approach to stationarity, such as the effective Hamiltonians whose spectral properties govern the loss of coherence among different symmetry sectors, and the signatures of AOSQMBS.

The rest of the paper is organized as follows: in Sec.\ref{Sec:2}, we start by reviewing the properties of bond and commutant algebras and their representations; we complete the aforesaid review by proposing a class of random brownian circuits for Lindblad evolution whose average operator autocorrelation dynamics is dictated by a frustration-free super-Hamiltonian having the commutant algebra as its kernel, extending an analogous theoretical proposal in~\cite{Moudgalya_2024} for closed quantum systems to the setting of markovian open quantum systems and paving the way for the investigation of slow relaxation modes in such systems. Next, in Sec.\ref{Sec:standard}, we provide simple applications of the algebraic formalism introduced in the preceeding section to common spin chains with discrete and $U(1)$ strong symmetries to acquantain the reader with its employ and to anticipate the conditions that we need to enforce in order to define models with exceptional stationary states besides typical ones. Finally, in Sec.\ref{Sec:4}, we introduce the notion of OSQMBS and present several examples of their occurrence in concrete Lindbladian systems, both in presence and in absence of an additional $U(1)$ symmetry, thereby investigating several dynamical signatures that result from their existence, ranging from approach to stationarity to the decay of off-diagonal coherences and the phenomenology of AOSQMBS.

\section{Commutant algebras and strong symmetries} \label{Sec:2}
\subsection{General methodological setting}
We consider in the following the time evolution of an open quantum system within the Markovian approximation, where it is described by the Lindblad equation:

\begin{equation} \label{Eq:Lindblad}
    \dot{\hat\rho} = -i[\hat H,\hat\rho]+\sum_j\gamma_j\left[\hat l_j \hat\rho \hat l^{\dag}_j-\frac{1}{2}\{\hat l^{\dag}_j\hat l_j,\hat\rho \}\right].
\end{equation}

We work under the further assumption that (i) the unitary part of the Lindblad dynamics is generated by a member of the family of Hamiltonians of the form $\hat H =\sum_{\alpha} J_{\alpha} \hat h_{\alpha}$, $\{\hat h_{\alpha}\}_{\alpha\in A}$ being a collection of local hermitian operators, and that (ii) the jump operators are hermitian, i.e., $\hat l^{\dag}_j=\hat l_j\,\,\forall j$. In such case, the dissipative term of the Lindblad dynamics can be rewritten as $\mathcal{D}(\hat\rho)=-\frac{1}{2}\sum_j \mathcal{O}^{\dag}_j( \mathcal{O}_j(\hat\rho))$, which means that it takes the form of a frustration-free superoperator, with $\mathcal{O}_j(\cdot):=[\hat l_j,\cdot]$ being the adjoint action of $\hat l_j$. Let us define the bond algebra $\mathcal{A}$ associated to a Lindblad system as the algebra generated by the identity operator, by the local terms $\{\hat h_{\alpha}\}$ in the unitary part of the evolution and by the jump operators $\{\hat l_j\}$. Its commutant $\mathcal{C}$ is then characterized as follows:
\begin{equation}
    \mathcal{C}=\{\hat O \in \mathcal{B}(\mathcal{H}):\, [\hat O, \hat g] = 0 \,\,\forall g \in \mathcal{A} \},
\end{equation}
i.e., it represents the algebra of all operators that commute with each local generator, and thus each element, of the bond algebra. Since $\mathcal{A}$ is a finite-dimensional unital subalgebra of $\mathcal{B}(\mathcal{H})$ closed under hermitian conjugation, it is a Von Neumann algebra and, moreover, $\mathcal{A}$ and $\mathcal{C}$ are centralizers of each other, by the double commutant theorem.
%Thus, the condition of stationarity for a state $\hat \rho_{ss}$ consists in being annihilated by the dissipative superoperator $\mathcal{D}$ of the Lindblad dynamics, which in turn is equivalent to stating that $\hat \rho_{ss}$ belongs to the commutant of the algebra generated by the set of jump operators $\hat l_j$. 

%Further assuming that the unitary part of the Lindblad dynamics is generated by a member of the family of Hamiltonians of the form $\hat H =\sum_{\alpha} J_{\alpha} \hat h_{\alpha}$, $\{\hat h_{\alpha}\}_{\alpha\in A}$ being a collection of (typically local) hermitian operators, then a stationary state $\hat \rho_{ss}$ of the Lindblad dynamics belongs the commutant algebra generated by both the set of jump operators $\{\hat l_j\}_{j\in J}$ and the set of local hamiltonian terms $\{\hat h_{\alpha}\}_{\alpha\in A}$.

The theory of representations of the bond algebra $\mathcal{A}$ and its commutant $\mathcal{C}$ states that the Hilbert space can be decomposed as a direct sum of tensor products of irreducible representations (irreps) of $\mathcal{A}$ and $\mathcal{C}$, namely:
\begin{align} \label{Eq:irreps}
\mathcal{H}=\bigoplus_{\lambda}\left(\mathcal{H}_{\lambda}^{(\mathcal{A})} \otimes \mathcal{H}_{\lambda}^{(\mathcal{C})} \right),
\end{align}
where $\mathcal{H}_{\lambda}^{(\mathcal{A})}$ and $ \mathcal{H}_{\lambda}^{(\mathcal{C})}$ are $D_{\lambda}-$ and $d_{\lambda}-$dimensional irreps of $\mathcal{A}$ and $\mathcal{C}$, respectively. Here, the set of tensor products of irreps, as well as the irreps' dimensions $d_{\lambda}$ and $D_{\lambda}$, labeled by $\lambda$, depends on the choice of the algebra $\mathcal{A}$. Furthermore, Eq.~\eqref{Eq:irreps} allows to identify Krylov subspaces corresponding to $\mathcal{A}$, i.e., $\mathcal{A}$-invariant subspaces that the bond algebra acts irreducibly upon: these are simply the subspaces $\mathcal{H}_{\lambda}^{(\mathcal{A})}$, each appearing with a degeneracy corresponding to the dimension of $\mathcal{H}_{\lambda}^{(\mathcal{C})}$, namely $d_{\lambda}$. Since for each block $\lambda$ there exist $d_{\lambda}$ $D_{\lambda}$-dimensional Krylov subspaces, the total number of Krylov subspaces is given by $N_K = \sum_{\lambda} d_{\lambda}$.

In order to understand the interplay between Lindblad dynamics and the representation theory of bond and commutant algebras, we observe that, by defining a basis $\mathcal{B}_{\lambda}=\{\ket{\lambda,m,a}:\, m=1,\dots, d_{\lambda},\, a=1,\dots,D_{\lambda}\}$ for each sector of the Hilbert space labeled by $\lambda$, the projectors $\hat\Pi_{m,m}^{\lambda}=\sum_a \ket{\lambda,m,a}\bra{\lambda,m,a}$ and the intertwine operators $\hat\Pi^{\lambda}_{m,m'}\sum_a \ket{\lambda,m,a}\bra{\lambda,m',a}$, with $m\neq m'$, span the commutant of $\mathcal{A}$ and form a complete set of the eigenspace with zero eigenvalue for a generic Liouvillian constructed out of the generators of $\mathcal{A}$. Thus, the expression of the stationary state resulting from the evolution from a given initial state $\hat\rho_0$ is determined solely by the overlap of $\rho_0$ with the conserved operators $\hat\Pi^{\lambda}_{m,m'}$ and is given by:
\begin{equation}\label{Eq:ss_comm}
    \hat\rho_{ss} = \sum_{\lambda,m,m'} Tr[\Pi^{\lambda}_{m',m}\hat\rho_0] \frac{\hat\Pi^{\lambda}_{m,m'}}{D_{\lambda}}.
\end{equation}
Additional off-diagonal coherences among sectors identified by different values of $\lambda$ are in principle, but are expected to be a nongeneric feature of the Lindblad evolution with arbitrarily chosen Hamiltonian couplings $J_{\alpha}$ and decay rates $\gamma_j$ and to appear only in presence of additional structure in the Lindbladian.
We will discuss an example of such behavior in section~\ref{Sec:4}, where we will discuss bond algebras featuring singlets, i.e., simultaneous eigenstates of all their local generators.

The commutant algebra framework allows as well for the identification of a complete set of conserved quantities that uniquely label the Krylov subspaces within the full Hilbert space. We identify them by examining the structure of what we will denote shortly as strong-symmetry group of $\mathcal{A}$. We remind that a \textit{strong symmetry} of the Lindblad dynamics generated by the bond algebra $\mathcal{A}$ is defined to be a unitary operator $\hat S$ satisfying $[\hat S, \hat H]=[\hat S, \hat l_j]=0\,\,\forall j$. Such operators identify subspaces in the operator Hilbert space that are left invariant under the Lindblad dynamics. 

The definition of strong symmetry implies that a strong symmetry is a unitary operator belonging to the commutant algebra $\mathcal{C}$ of the bond algebra $\mathcal{A}$. Thus, the strong symmetries associated to the Lindblad evolution with Hamiltonian bond terms $\{\hat h_{\alpha}\}_{\alpha\in A}$ and jump operators $\{\hat l_j\}_{j\in J}$ form a group, which we dub strong-symmetry group, that is isomorphic to $\bigoplus_{\lambda} U(d_{\lambda})$ and each such strong symmetry $\hat S$ can be represented as:
\begin{equation}
    \hat S = \bigoplus_{\lambda}\left(\mathbb{I}_{D_{\lambda}}^{(\mathcal{A})} \otimes M^{(\mathcal{C})}_{d_{\lambda}}(\hat S) \right),
\end{equation}
where $M^{(\mathcal{C})}_{d_{\lambda}}(\hat S)$ is a $d_{\lambda}\times d_{\lambda}$ unitary matrix representing its action within $\mathcal{H}_{\lambda}^{(\mathcal{C})}$. 

In order to quantify the number of mutually-commuting strong symmetries, we observe that the dimension of a maximal abelian subalgebra of the Lie algebra $u(d_{\lambda})$ of the Lie group $U(d_{\lambda})$ is $d_{\lambda}$, thus resulting in $\sum_{\lambda} d_{\lambda}$ mutually-commuting strong symmetries. The latter are seen to take the form:
\begin{equation} \label{Eq:strong}
    \hat S^{(\lambda,n)} =  \mathbb{I}_{D_{\lambda}}^{(\mathcal{A})} \otimes D^{(\mathcal{C})}_{n,d_{\lambda}}, \quad n=1,\dots , d_{\lambda},
\end{equation}
where we introduced the $d_{\lambda}\times d_{\lambda}$ matrix $[D^{(\mathcal{C})}_{n,d_{\lambda}}]_{ij}= \delta_{ij}[\delta_{jn} e^{i\theta_{n,\lambda}}+1-\delta_{jn}]$, for any choice of the value of $\theta_{n,\lambda}\neq 0$. 
In this way, we have identified $\sum_{\lambda} d_{\lambda}$ subspaces, corresponding to a choice of Krylov subspaces of $\mathcal{A}$ within the Hilbert space $\mathcal{H}$, that are invariant and irreducible under the action of $\mathcal{A}$ and can be uniquely labelled by the set of eigenvalues of the strong symmetries in Eq.~\eqref{Eq:strong}. 

%Alternatively, one can more simply derive the above result by exploiting the isomorphism $\mathcal{B}(\mathcal{H})\cong \mathcal{H} \otimes \mathcal{H}^*$ for the operator Hilbert space $\mathcal{B}(\mathcal{H})$ in conjunction with relation~\eqref{Eq:irreps}. The following diagonal operator subspaces:
%\begin{align}
%    W_{\lambda,n}:=span\{\ket{\psi}\bra{\phi}: \ket\psi,\,\ket{\phi}\in \mathcal{H}_{\lambda}^{(\mathcal{A})} \otimes \mathcal{H}_{\lambda}^{(n,\mathcal{C})} \},
%\end{align}
%where we have decomposed $\mathcal{H}_{\lambda}^{(\mathcal{C})}$ as $\bigoplus_{n=1}^{d_{\lambda}}\mathcal{H}_{\lambda}^{(n,\mathcal{C})}$ into its one-dimensional subspaces spanned by the eigenvectors of $\hat S^{(\lambda,n)}$, are such that the action of the algebra $\mathcal{A}$ is irreducible within them. Since operators with non-vanishing trace can only be found in such subspaces $W_{\lambda,n}$, we obtain one stationary state for each such subspace, thus leading to the result $N_{st}= \sum_{\lambda}d_{\lambda}$ for the number of stationary states $N_{st}$.

\subsection{Brownian circuit interpretation}\label{Subsec:brownian}

The commutant $\mathcal{C}$ of the algebra $\mathcal{A}$ generated by the family of chosen hamiltonian bond terms and jump operators can be alternatively described as the ground-state space of the frustration-free local super-Hamiltonian~\cite{Moudgalya_2024}:
\begin{equation}\label{Eq:diss_superham}
    \mathcal{P} = \sum_{\alpha} \mathcal{L}^{\dag}_{\alpha} \mathcal{L}_{\alpha}+\sum_j \mathcal{L}^{\dag}_{j} \mathcal{L}_{j},
\end{equation}
where $\mathcal{L}_{\alpha}=[\hat h_{\alpha},\cdot]$ and $\mathcal{L}_{j}=[\hat l_j , \cdot]$. As in the case of unitary evolution of closed quantum many-body systems, the super-Hamiltonian was interpreted as the generator of effective ensemble-averaged evolution in random Brownian circuits associated to a given bond algebra~\cite{Moudgalya_2024}, it is natural to ask whether the super-Hamiltonian in Eq.~\eqref{Eq:diss_superham} admits a natural interpretation in the framework of open quantum system dynamics along similar lines.

We show here that the super-hamiltonian in Eq.~\eqref{Eq:diss_superham} appears naturally as the generator of operator relaxation within Lindbladian brownian circuits with random Hamiltonian coefficients and deterministic decay rates. To this end, let us consider a Lindbladian brownian circuit, i.e., a quantum circuit for the finite-step time evolution of the operators at times $t_j=\epsilon j$ according to the rule:
\begin{equation}
\begin{split}
    &\hat O(t_{j+1})=e^{\epsilon \mathcal{L}_j^{\dag}}\hat O(t_j),\\
    &\mathcal{L}^{\dag}_j[\cdot]= i\sum_{\alpha\in A} g_{\alpha}^{(j)}[\hat h_{\alpha},\cdot]-\frac{1}{2}\sum_{l\in J} \gamma_l [\hat l_l,[\hat l_l,\cdot]],
    \end{split}
\end{equation}
where the decay rates $\{\gamma_l\}_{l\in J}$ are fixed, while the Hamiltonian's coefficients $\{g_{\alpha}^{(j)} \}_{\alpha\in A}$ are randomly distruibuted according to a gaussian probability density function satisfying:
\begin{equation}
    \langle g_{\alpha}^{(j)} \rangle=0, \quad \langle g_{\alpha}^{(j)} g_{\beta}^{(l)} \rangle =\delta_{\alpha\beta}\delta_{jl} \frac{2k_{\alpha}}{\epsilon}, \,\, k_{\alpha}>0.
\end{equation}

Upon averaging over the Hamiltonian's coefficients, indicated by an upper bar, the time evolution of the infinite-temperature autocorrelation function of the operator $\hat O$, defined by the relation $\langle \hat O  (t_j) \hat O \rangle= Tr[\hat O(t_j) \hat O]/D$, $D$ being the Hilbert space dimension, obeys the relation:
\begin{equation}
    \begin{split}
        & \langle \overline{\hat O  (t_j+\epsilon)}\hat O\rangle= \langle\overline{\hat O  (t_j)}\hat O\rangle-\epsilon \langle \mathcal{D}_{eff}(\overline{\hat O(t_j)}) \hat O\rangle,\\
        & \mathcal{D}_{eff}(\cdot):=\frac{1}{2}\sum_{l\in J} \gamma_{l} [\hat l_l ,[\hat l_l,\cdot]] +\sum_{\alpha\in A} k_{\alpha} [\hat h_{\alpha},[\hat h_{\alpha},\cdot]].
    \end{split}
\end{equation}
Upon taking the continuum limit, the function $\langle\overline{ \hat O(t)} \hat O \rangle$ evolves as:
\begin{align}\label{Eq:op_relax}
   \langle \overline{\hat O(t)}\hat O \rangle = \frac{Tr[ e^{-t \mathcal{D}_{eff}}(\hat O)\,\, \hat O]}{D}.  
\end{align}

As an application of such a picture, we show that the Mazur bound~\cite{Moudgalya_2024,Moudgalya_2022,Mazur_1969,Dhar_2021,Li_2023} on infinite-temperature autocorrelation functions in Lindblad quantum systems appears as the stationary value of the evolution in Eq.~\eqref{Eq:op_relax}. By expanding the operator $\hat O$ in the basis of eigenstates $\{\hat \lambda_{\alpha} \}_{\alpha=1}^{D^2}$ with eigenenergies $\{\epsilon_{\alpha}\}_{\alpha=1}^{D^2}$ of the super-Hamiltonian $\mathcal{D}_{eff}$ and taking the long-time limit $t\rightarrow +\infty$, only the eigenoperators of $\mathcal{D}_{eff}$ with eigenvalue zero contribute to the infinite-time infinite-temperature autocorrelation function, which reads:
\begin{equation}\label{Eq:Mazur}
    \lim_{t \to +\infty} \frac{Tr[ e^{-t \mathcal{D}_{eff}}(\hat O)\,\, \hat O]}{D} = \frac{1}{D}\sum_{\alpha} \delta_{\epsilon_{\alpha},0} \Bigl|Tr[\hat\lambda_{\alpha}^{\dag} \hat O]\Bigr|^2. 
\end{equation}
The r.h.s. of Eq.~\eqref{Eq:Mazur} coincides with the Mazur bound on the infinite-temperature autocorrelation function of the operator $\hat O$ in presence of the set of linearly-independent conserved quantities $\{\hat\lambda_{\alpha}:\,\alpha \in\{1,\,\dots,D^2\}\,s.t.\,\epsilon_{\alpha} = 0\}$, which form an orthonormal basis of the commutant of the chosen bond algebra.

\section{Application to spin chains with $\mathbb{Z}_2$ and $U(1)$ symmetries}\label{Sec:standard}
In the following, we consider bond algebras appearing in typical spin systems and apply the results above to characterize the convex set of their stationary states and possibly estimate their expressions. We start with the following lemma.

\textbf{Lemma}: \textit{let us consider a bond algebra $\mathcal{A}$ such that its commutant $\mathcal{C}$ is abelian. Then, the family of stationary states is given by all convex combinations of the trace-normalized projectors onto the Krylov subspaces of $\mathcal{A}$; each such state represents the unique stationary state for every initial condition of the Lindblad dynamics that lies entirely within the corresponding Krylov subspace. Moreover, the number of Krylov subspaces equals the dimension of the commutant (as a vector space).}

\textbf{Proof}: since the case commutant is abelian, it can only have one-dimensional irreducible representations, i.e., $d_{\lambda}=1\,\,\forall \lambda$. Moreover, whenever the initial state lies entirely within a Krylov subspace $\mathcal{K}_{\lambda}$ of $\mathcal{A}$, it converges to a unique stationary state, given by the trace-normalized projector onto such Krylov subspace, i.e., $\hat\rho_{ss,\lambda}=\hat\Pi^{\lambda}/D_{\lambda}$~\cite{Li_2023,Li_2025}. Thus, the family of stationary states is formed by all convex combinations of the operators of the form $\{\hat\rho_{ss,\lambda} \}_{\lambda\in S}$, $S$ being the set of different allowed values of $\lambda$. Additionally, the number of Krylov sectors, which we indicate as $|S|$, satisfies:
\begin{equation}
    dim(\mathcal{C})=\sum_{\lambda\in S} d_{\lambda}^2=\sum_{\lambda\in S} 1=|S|.
\end{equation}
Hence, the number of linearly-independent stationary states, one for each Krylov subspace of the bond algebra $\mathcal{A}$, equals the dimension of the commutant in the case of abelian commutants, QED.

Keeping this result in mind, we construct the first example by taking an algebra of the form $\mathcal{A} = \langle\langle \{\hat \sigma_j^x \hat \sigma_{j+1}^x\},\{ \hat \sigma_j^z\},\{ \hat\sigma_j^x\} \rangle \rangle$, where $\hat\sigma_j^{\alpha}$, $\alpha=x,y,z$, denotes the Pauli matrix with component $\alpha$ at site $j$. Since $\hat\sigma_x\hat\sigma_z\propto \sigma_y$, it is easy to see that $\mathcal{A}$ coincides with the full operator space on the Hilbert space of the spin system, thus resulting in the trivial commutant $\mathcal{C}=\langle\langle \{\mathds{\hat 1}\} \rangle \rangle$, together with the unique infinite-temperature stationary state $\hat\rho_{ss}=\mathds{\hat 1}/2^L$, $L$ being the size of the system.

Further, we consider the bond algebra $\mathcal{A}=\langle\langle \{\hat \sigma_j^x \hat \sigma_{j+1}^x\},\{ \hat \sigma_j^z\} \rangle \rangle$, whose commutant can be shown to take the form $\mathcal{C}=\langle\langle  \prod_j \hat\sigma_j^z  \rangle\rangle$. Since $dim(\mathcal{C})=2$ in this case, we obtain that there exist two Krylov subspaces, each one hosting a unique stationary state. The Krylov subspaces in such case can be identified with the parity sectors of the Hilbert space, labeled by the eigenvalue of the parity operator $\prod_j \hat \sigma_j^z$, and the stationary states residing in each of them are the (trace-normalized) projectors $\hat\Pi_{z}^{\pm}$ onto fixed-parity sectors:
\begin{equation}
  \hat \hat\rho_{ss,\pm}=\frac{\Pi_{\pm}^z}{Tr[\hat \Pi_{\pm}^z]} = \frac{1}{2^L}\left(\mathds{\hat 1}\pm \prod_j \hat \sigma_j^z\right).  
\end{equation}

We continue our discussion with the bond algebra $\mathcal{A}=\langle\langle \{\hat \sigma_j^x \hat \sigma_{j+1}^x\},\{ \hat \sigma_j^z \hat \sigma_{j+1}^z\}\rangle\rangle$, whose commutant reads $\mathcal{C}=\langle\langle \prod_j\hat \sigma_j^x,\prod_j \hat \sigma_j^z\rangle\rangle$. For simplicity, we set $L$ to be even, so that $\mathcal{C}$ is abelian. Now, an orthogonal basis for the commutant is given by $\{1,\prod_j\hat \sigma_j^x,\prod_j\hat \sigma_j^z,\prod_j\hat \sigma_j^y \}$ (since $\left(\prod_j\hat \sigma_j^x\right)\left(\prod_j\hat \sigma_j^z\right)=(-i)^{L}\prod_j\hat \sigma_j^y$). Hence, we deduce that $dim(\mathcal{C}) = 4$, with $4$ mutually orthogonal stationary states being given by:
\begin{equation}
\begin{split}
   &\hat\rho_{ss;\alpha,\beta}= \frac{\Pi_{\alpha}^z \Pi_{\beta}^x}{Tr[\Pi_{\alpha}^z \Pi_{\beta}^x]}=\frac{1}{2^L}\left(\mathds{\hat 1}+\alpha \prod_j \hat \sigma_j^z\right)\left(\mathds{\hat 1}+\beta \prod_j \hat \sigma_j^x\right),\\
   & \alpha, \,\beta=\pm,
   \end{split}
    \end{equation} 
which correspond to the available choices of the eigenvalues of $\prod_j\hat \sigma_j^x$ and $\prod_j\hat \sigma_j^z$, that label the Krylov subspaces of the given bond algebra $\mathcal{A}$.

Finally, we mention the bond algebra $\mathcal{A}=\langle\langle \{\hat \sigma_j^x \hat \sigma_{j+1}^x+\hat\sigma_j^y \hat\sigma_{j+1}^y\},\{ \hat \sigma_j^z \}\rangle\rangle$, whose commutant takes the form $\mathcal{C}=\langle\langle  \hat \sigma_{tot}^z\rangle\rangle$. Since the algebra is spanned by the basis $\{\mathds{\hat 1}, \hat\sigma^z_{tot},\dots,(\hat\sigma^z_{tot})^L\}$, we get the relation $dim(\mathcal{C}) = L+1$. The stationary states that identify the Krylov subspaces of $\mathcal{A}$ correspond in this case to the trace-normalized projectors onto each of the magnetization sectors of the Hilbert space, labeled by the $L+1$ eigenvalues $\{-L,-L+2,\dots,L-2,L\}$ of the total $z$-magnetization operator $\hat\sigma_{tot}^z=\sum_j \hat \sigma_j^z$. 

Closing this section, we notice that the last bond algebra that we considered possesses two singlets, i.e., states belonging to a sector with $D_{\lambda}=1$, which are simultaneous eigenstates of all generators (the fully-polarized states $\ket{\uparrow\dots\uparrow},\ket{\downarrow\dots\downarrow}$). In order to define quantum many-body scars (QMBS) in an open setting, it would be desirable to promote projectors onto singlets to the unique generators of the commutant, by introducing bond terms that preserve the invariance of the singlets under the action of $\mathcal{A}$ while removing the additional $U(1)$ symmetry. This is motivated by the idea that OSQMBS shall be defined as projectors onto invariant one-dimensional Krylov subspaces that are not protected by any additional symmetry structure encoded in the symmetry algebra $\mathcal{C}$. We are thus ready to introduce the notion of OSQMBS in the commutant-algebra language in the next section and apply it to some concrete examples.
\section{QMBS in open quantum systems}\label{Sec:4}

\subsection{Definition and general properties}

Despite the many diverse perspectives and formalisms to tackle the phenomenon of quantum many-body scarring, the notion of QMBS is already well established and deeply discussed in the literature on weak-ergodicity breaking in closed quantum systems~\cite{Turner_2018,Turner_2018_2,Choi_2019,Shibata_2020,Moudgalya_2020,Moudgalya_2020_2,Moudgalya_2020_3,Iadecola_2019,Iadecola_2020,Schecter_2019,Gotta_2022}, while it remains a largely open question whether it is possible to characterize an analogous phenomenon in markovian open quantum systems described by the Lindblad equation. We propose to characterize a QMBS for markovian open quantum systems, which we named as OSQMBS, as a pure state, that is a stationary state and such that any initial state that is orthogonal to it converges to the infinite-temperature state in the orthogonal complement to the one-dimensional subspace spanned by the OSQMBS. Formally, we define a pure state $\hat \rho_{\psi}=\ket{\psi}\bra{\psi}$ to be an OSQMBS iff it satisfies $\mathcal{L}(\hat\rho_{\psi})=0$ and, for every initial state $\hat \rho(0)$ such that $Tr[\rho_{\psi}\rho(0)]=0$, one has:
\begin{equation}
    \lim_{t\rightarrow +\infty} e^{\mathcal{L}t}\rho(0)=\frac{\mathds{\hat 1}-\hat\rho_{\psi}}{d-1},
\end{equation}
$d$ being the Hilbert space dimension. According to such a definition, $\hat \rho_{\psi}$ can be interpreted as an exceptional state that does not evolve towards the infinite-temperature state, contrarily to every physical state orthogonal to it.

The key technical step towards the construction of models that display the aforesaid phenomenology is the construction of bond algebras $\mathcal{A}$ with a distinctive algebraic property: the corresponding commutant $\mathcal{C}$ is spanned solely by the identity operator and the projector onto a state $\ket{\psi}$, namely, $\mathcal{C}=\langle\langle \ket{\psi}\bra{\psi} \rangle\rangle$. In such a case, Eq.~\eqref{Eq:irreps} will be realized with two blocks, which we dub scar and thermal blocks, labeled by $\lambda=\lambda_{sc},\lambda_{th}$ and naturally identified with the linear span of the state $\ket{\psi}$ and its orthogonal complement. Since the commutant is abelian, its irreducible representations are one-dimensional, thus giving $d_{\lambda_{th}}=d_{\lambda_{sc}}=1$. On the other hand, the algebra $\mathcal{A}$ leaves the scar block invariant, namely, $\ket{\psi}$ is a singlet of $\mathcal{A}$, thus leading to $D_{\lambda_{sc}}=1$, while the action of $\mathcal{A}$ within the thermal block is irreducible; from the relation $\sum_{\lambda} D_{\lambda} d_{\lambda}=dim(\mathcal{H})$ it is then possible to obtain that $D_{\lambda_{th}}=dim(\mathcal{H})-1$.

The implications of the above considerations on the structure of the stationary states of the corresponding Lindblad evolution can be drawn directly from Eq.~\eqref{Eq:ss_comm} and lead to the statement that then the stationary-state manifold is fully generated by the convex combinations of the projector $\ket{\psi}\bra{\psi}$ onto the scar state and of the projector onto the orthogonal complement $(span\{\ket{\psi}\})^{\perp}$, which we dub the thermal subspace. Formally, the generic expression for a stationary state $\hat\rho_{ss}(p)$ of such a system will be:
\begin{equation}
    \hat \rho_{ss}(p) = p \ket{\psi}\bra{\psi} +(1-p) \frac{\mathds{\hat 1}-\ket{\psi}\bra{\psi}}{d-1},\quad 0\leq p \leq 1,
\end{equation}
where the value of $p=Tr[\ket{\psi}\bra{\psi}\hat\rho(0)]$, given by the overlap of the initial condition $\hat\rho(0)$ with the OSQMBS $\ket{\psi}\bra{\psi}$, depends on the initial condition itself of the time evolution.

The setup presented above, in which the Hilbert space is the direct sum of a one-dimensional subspace, spanned by a singlet of the algebra $\mathcal{A}$, and its orthogonal complement, the latter forming an irrep of $\mathcal{A}$, allows to keep track of the fate of off-diagonal coherences between the two aforementioned sectors, contrarily to what happens in the most generic case. The latter can be quantified by computing the time evolution of the weight of a generic time-evolving density matrix over coherences of the form $\ket{\psi}\bra{\varphi}$, with $\ket{\varphi}\in \, span(\{\ket{\psi} \})^{\perp}$. More formally, we show that:
\begin{equation}\label{Eq:coh_off}
    \begin{split}
    &\frac{d}{dt} ||\hat\Pi_s e^{\mathcal{L}t}(\hat\rho_0) \hat\Pi_{th}||^2 =\\
    &=-Tr\left[e^{\mathcal{L}t}\left( \hat\Pi_{s} \hat \rho_0 \hat\Pi_{th} \right)\hat H_2 e^{\mathcal{L}t}\left( \hat\Pi_{th} \hat \rho_0 \hat\Pi_{s} \right)\right],
    \end{split}
\end{equation}
where $\hat\Pi_s= \ket{\psi}\bra{\psi}$ and $\hat\Pi_{th}=1-\hat\Pi_{s}$ (but the same conclusion holds even if $\hat\Pi_{th}$ is the projector onto a Krylov subspace associated to the action of $\mathcal{A}$ on the Hilbert space $\mathcal{H}$ in presence of a more refined symmetry resolution of $\mathcal{H}$, as we will see below). A derivation and the form of $\hat H_2 \geq 0$ can be found in the Supplementary Material (SM). Thus, depending on the spectral properties of $\hat H_2 $ in the subspace $\hat\Pi_{th}$, which is invariant under the action of $\hat H_2$, the decay of the off-diagonal coherences between the subspace identified by $\hat \Pi_s$ and its orthogonal complement can change dramatically. In particular, if $\hat\Pi_{th}\hat H_2 \hat\Pi_{th}\geq g>0$, meaning that $\hat H_2$ is strictly positive over the range of $\hat\Pi_{th}$, then the aforesaid coherences are bound to decay at least exponentially fast.

All of the above considerations can be readily generalized to the alternative paradigmatic case of commutants of the form $\mathcal{C}=\langle\langle \{\ket{\psi_n}\bra{\psi_m}\}_{0\leq m,n\leq N} \rangle \rangle$, i.e., coinciding with the operator algebra associated to a subspace spanned by a tower of $N+1$ states $\{\ket{\psi_n} \}_{n=0}^N$ (typically with $N\sim O(L)$, $L$ being the size of the system), that are orthonormal and iteratively generated through a spectrum-generating-algebra relation starting from a reference "vacuum" state~\cite{Serbyn_2021,Moudgalya_2022_review,Chandran_2023}. In such case, the Hilbert space decouples in a thermal block, labelled by $\lambda_{th}$, and in a scar block, denoted with the label $\lambda_{sc}$. While the action of the commutant is trivial within the thermal block and irreducible within the scar block, resulting in $d_{\lambda_{th}}=1$ and $d_{\lambda_{sc}}=N+1$, the action of the corresponding bond-algebra $\mathcal{A}$ within the thermal block is irreducible and within the scar block is trivial, thus giving $D_{\lambda_{th}}=D-N-1$ and $D_{\lambda_{sc}}=1$. 

As a final consideration, we notice that a typical modification of the above scenario occurs in presence of an extensive $U(1)$ charge $\hat Q$ that "lifts" the degeneracy among the states in the tower $\{\ket{\psi_n} \}_{n=0}^N$, meaning that $\hat Q\ket{\psi_n} = n\ket{\psi_n},\, 0\leq n \leq N$. If one adds the extensive operator $\hat Q$ to the set of generators of the bond algebra $\mathcal{A}$, the commutant takes the form $\mathcal{C}=\langle\langle \{\ket{\psi_n}\bra{\psi_n}\}_{0\leq n\leq N} \rangle \rangle$, and the Hilbert space splits into the direct sum of $N+1$ one-dimensional Krylov subspaces, each spanned by one of the states $\ket{\psi_n}$ belonging to the tower and satisfying $d_{\lambda_{sc},n}=D_{\lambda_{sc,n}}=1\,\,\forall n$, with their orthogonal complement, which satisfies $D_{\lambda_{th}}=D-N-1$ and $D_{\lambda_{sc}}=1$ as before. 

In the additional instance where $\hat Q$ is a symmetry itself, so that the commutant becomes $\mathcal{C} =\langle\langle \{\ket{\psi_n}\bra{\psi_n}\}_{0\leq n\leq N}, \hat Q \rangle \rangle$, the thermal subspace is further resolved into sectors of fixed value of the given $U(1)$ charge $\hat Q$, thus resulting in a nontrivial encoding of the interplay of $U(1)$ symmetry and scar symmetry into the Hilbert-space representation. We will consider more in detail such a case in Subsection~\ref{Subsec:tower}.  

%\textcolor{red}{In order to realize such a structure, we refer to the work in \href{https://arxiv.org/abs/2209.03377}{Ref.}, where commutant algebras generated by projectors onto QMBS (and eventually ket/bra operators between different QMBS eigenstates) have been constructed. Namely, it is possible to construct local algebras $\mathcal{A}_{scar}$ such that their commutant reads $C_{scar}=\langle\langle\{\ket{\psi_m}\bra{\psi_n}\}_{m,n\in S}\rangle\rangle$, $S$ being an index set. A major instance of such construction is presented in Lemma IV.1 of \href{https://arxiv.org/abs/2209.03377}{Ref.}, which ensures that it is always possible to find an algebra generated by strictly local hermitian operators such that its commutant is generated by all ket-bra operators of states forming a basis of a subspace of the form $\mathcal{T}=\{\ket{\psi}: \hat P_j \ket{\psi}=0\,\,\forall j\}$, $P_j$ being strictly local projectors.
%A major implication of such lemma is the case of an injective MPS, which is known to be the unique state in the common kernel of some projector operators, whose range depends on the bond dimension. It is thus possible to realize a local algebra $\mathcal{A}_{\psi}$ such that $\mathcal{C}_{\psi}=\langle\langle\ket{\psi}\bra{\psi}\rangle\rangle$. Similar constructions have been verified for towers of states (see Eqs. $(22,26,30,34)$ of \href{https://arxiv.org/abs/2209.03377}{Ref.}).}

\subsection{Examples of isolated OSQMBS}

As a first example of lindbladian systems with isolated OSQMBS~\cite{Bocini_2024}, we introduce the algebra $\mathcal{A}=\langle\langle \{\hat \sigma_j^x \hat \sigma_{j+1}^x+\hat\sigma_j^y \hat\sigma_{j+1}^y\}, \{\hat \sigma_j^z\} , \{\hat\sigma_j^x (1-\hat \sigma_{j+1}^z)\}, \{(1-\hat\sigma_j^z)\hat \sigma_{j+1}^x \} \rangle\rangle$~\cite{Marche_2024}. The commutant of $\mathcal{A}$ is $\mathcal{C}=\langle\langle \ket{F,\uparrow} \bra{F,\uparrow} \rangle\rangle$, i.e., it is generated by the projector onto the fully upward-polarized ferromagnetic state:
\begin{equation}
    \ket{F,\uparrow}:=\ket{\uparrow\dots\uparrow},
\end{equation}
 and by the projector onto the orthogonal complement $\left(span\{ \ket{F,\uparrow} \bra{F,\uparrow} \}\right)^{\perp}$. Formally, the Hilbert space decomposes thus as:
\begin{align} 
\mathcal{H}=\left(\mathcal{H}_{\lambda_{th}}^{(\mathcal{A})} \otimes \mathcal{H}_{\lambda_{th}}^{(\mathcal{C})} \right)\bigoplus \left(\mathcal{H}_{\lambda_{sc}}^{(\mathcal{A})} \otimes \mathcal{H}_{\lambda_{sc}}^{(\mathcal{C})} \right),
\end{align}
where $D_{\lambda_{th}}=D-D_{\lambda_{sc}}=2^L-1$. In order to obtain $\mathcal{C}$ for the aforesaid $\mathcal{A}$, one shows that the algebra $\mathcal{A}$ acts irreducibly on the orthogonal complement of the one-dimensional subspace spanned by the scar state $\ket{F,\uparrow}$, namely, that $\ket{\psi_{\alpha}}\bra{\psi_{\beta}}\in \mathcal{A}$ for all elements $\ket{\psi_{\alpha}},\,\ket{\psi_{\beta}}$ of a $2^L-1$-dimensional basis of $\left(span\{\ket{F,\uparrow}\}\right)^{\perp}$, as carried out in the SM. In light of the general framework developed above, generic lindbladians that feature the generators of the algebra $\mathcal{A}$ as local terms within the Hamiltonian and the jump operators are going to possess a typical stationary state $\hat\Pi_{th}:=\hat{\Pi}_{span\{\ket{F,\uparrow}\}^{\perp}}$ and the OSQMBS $\ket{F,\uparrow}\bra{F,\uparrow}$. 

Concerning the behavior of coherences between the OSQMBS and its orthogonal subspace in Eq.~\eqref{Eq:coh_off}, it is easy to estimate what happens when the jump operators satisfy $\hat l_j = \hat \sigma_j^z\,\,\forall j$: in such case, indeed, $\hat H_2 =4 \sum_j \gamma_j (1-\hat\sigma_j^z)/2$, which satisfies $\hat\Pi_{th}\hat H_2 \hat\Pi_{th}\geq 4 (\min_j \gamma_j)$, and thus the off-diagonal coherences are bound to decay at least exponentially fast.

Similarly, in order to show an example of algebra $\mathcal{A}$ that features two isolated OSQMBS as exceptional stationary states, let us consider the algebra $\mathcal{A}=\langle\langle\{\hat \sigma_j^x \hat \sigma_{j+1}^x+\hat\sigma_j^y \hat\sigma_{j+1}^y\}, \{\hat \sigma_j^z\} ,\{ \hat\sigma_j^+ \hat \sigma_{j+1}^+ \hat \sigma_{j+2}^- +h.c. \} \rangle\rangle$. As before, if not for the last bond term, this algebra would coincide with the $U(1)$-conserving algebra described in section~\ref{Sec:standard}. However, the last set of generators violates magnetization conservation, while annihilating the two fully-polarized states $\ket{F,\uparrow}$ and:
\begin{equation}
    \ket{F,\downarrow}:=\ket{\downarrow\dots\downarrow}.
\end{equation}
 Thus, we expect the commutant to be $\mathcal{C}=\langle\langle\ket{F,\uparrow}\bra{F,\uparrow},\ket{F,\downarrow}\bra{F,\downarrow}  \rangle\rangle$. One can once again formally derive this result by proving that the algebra $\mathcal{A}$ acts irreducibly in the orthogonal complement of $W_{scar}:=span\{ \ket{F,\uparrow},\ket{F,\downarrow} \}$, i.e., it contains the ket-bra operator $\ket{\psi_{\alpha}}\bra{\psi_{\beta}}$ formed by any two basis states $\ket{\psi_{\alpha}},\, \ket{\psi_{\beta}}$ for such $2^L-2$-dimensional subspace, as shown in the SM. As a result, the Hilbert space will break into the direct sum:
\begin{align} 
\mathcal{H}=\left(\mathcal{H}_{\lambda_{th}}^{(\mathcal{A})} \otimes \mathcal{H}_{\lambda_{th}}^{(\mathcal{C})} \right)\bigoplus \left[\bigoplus_{\alpha=\uparrow,\downarrow}\left(\mathcal{H}_{\lambda_{sc,\alpha}}^{(\mathcal{A})} \otimes \mathcal{H}_{\lambda_{sc,\alpha}}^{(\mathcal{C})} \right)\right],
\end{align}
where $D_{\lambda_{sc,\uparrow}}=D_{\lambda_{sc,\downarrow}}=d_{\lambda_{sc,\uparrow}}=d_{\lambda_{sc,\downarrow}}=d_{\lambda_{th}}=1$ and $D_{\lambda_{th}}=D-\sum_{\alpha=\uparrow,\downarrow}D_{\lambda_{sc,\alpha}}=2^L - 2$.
Analogously to what assessed before, generic lindbladians that feature the generators of the algebra $\mathcal{A}$ as local terms within the Hamiltonian and the jump operators are going to possess a typical stationary state $\hat{\Pi}_{(span\{\ket{F,\uparrow},\ket{F,\downarrow}\})^{\perp}}/(2^L-2)$ and two OSQMBS stationary states $\ket{F,\uparrow}\bra{F,\uparrow}$, $\ket{F,\downarrow}\bra{F,\downarrow}$. 

\begin{figure}
\centering
\includegraphics[width=0.5\textwidth]{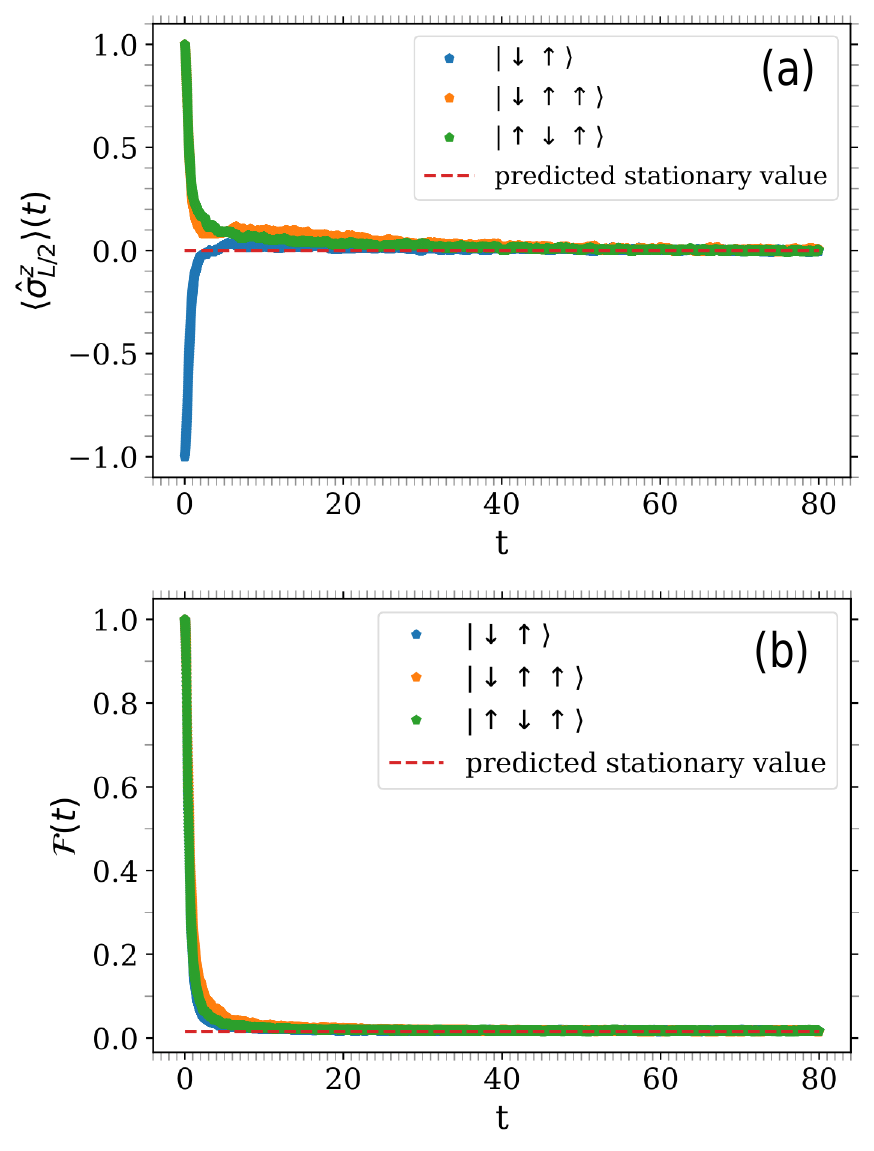}
\caption{(a) Expectation value $\langle \hat \sigma_{L/2}^z \rangle (t) =Tr[\hat \sigma_{L/2}^z \hat\rho(t)]$ as a function of time $t$ for parameter values $(J,D,\gamma) = (0.5, 1.2, 1.0)$ at size $L=6$ and with initial states obtained by the periodic repetition of the patterns indicated in the legend until exhaustion of the number of sites $L$. (b) Fidelity $\mathcal{F} (t) =Tr[\ket{\psi(0)}\bra{\psi(0)} \hat\rho(t)]$ as a function of time $t$ for the same parameter values, system size $L$ and initial states as in (a).}
\label{Fig:isolated}
\end{figure}

We corroborate these theoretical observations by performing numerical simulations with a simple first-order quantum trajectories' algorithm\cite{Daley_2014}, implemented within the open-source Python package QuSpin\cite{Quspin}. We consider a Lindbladian realized from the assigned algebra $\mathcal{A}$, which we choose to be of the form:
\begin{equation}
    \begin{split}
        &\mathcal{L} = -i[\hat H, \cdot]-\frac{\gamma}{2}\sum_{j=1}^L [\hat l_j,[\hat l_j,\cdot]],\\
        & \hat H = -J\sum_{j=1}^{L-1}\left[\hat\sigma_j^x \hat\sigma_{j+1}^x +\hat\sigma_j^y \hat\sigma_{j+1}^y \right] +D \sum_{j=1}^{L-2}\left[\hat\sigma_j^+ \hat\sigma_{j+1}^+ \hat\sigma_{j+2}^{-}+h.c.\right],\\
        & \hat l_j = \hat\sigma_j^z,
    \end{split}
\end{equation}
where we fixed open boundary conditions (OBC). In Fig.~\ref{Fig:isolated}, we show how, for a generic choice of parameters $(J,D,\gamma)$, we observe indeed that an arbitrary initial state of the dynamics that is orthogonal to both $\ket{F,\uparrow}$ and $\ket{F,\downarrow}$ reproduces the properties of the typical stationary state $\hat{\Pi}_{(span\{\ket{F,\uparrow},\ket{F,\downarrow}\})^{\perp}}/(2^L-2)= (1-\ket{F,\uparrow}\bra{F,\uparrow}-\ket{F,\downarrow}\bra{F,\downarrow})/(2^L - 2)$ in the large-time limit. More specifically, we monitor both the time evolution of $\langle \hat\sigma_{L/2}^z \rangle(t):= Tr[\hat\sigma^z_{L/2} \hat\rho(t)]$ and of the fidelity $\mathcal{F}(t):= Tr[\ket{\psi(0)}\bra{\psi(0)} \hat \rho(t)]$, where $\ket{\psi(0)}$ is the initial state, chosen to be orthogonal to $\ket{F,\uparrow}$ and $\ket{F,\downarrow}$, and $\hat\rho(t)$ is the density matrix of the system at time $t$; we observe that their large-time stationary values are compatible with the ones obtained from the aforesaid typical stationary state, which read $Tr[\hat\sigma^z_{L/2}(1-\ket{F,\uparrow}\bra{F,\uparrow}-\ket{F,\downarrow}\bra{F,\downarrow})]/(2^L-2)=0$ and $Tr[\ket{\psi(0)}\bra{\psi(0)}(1-\ket{F,\uparrow}\bra{F,\uparrow}-\ket{F,\downarrow}\bra{F,\downarrow})]/(2^L-2)=1/(2^L-2)$, respectively.

Finally, in order to characterize the behavior of the off-diagonal coherences, we assume once more the typical scenario of dephasing, namely $\hat l_j = \hat\sigma_j^z\,\,\forall j$. Then, when applying Eq.~\eqref{Eq:coh_off} to the subspaces identified respectively by the range of $\hat\Pi_{\sigma}=\ket{F,\sigma}\bra{F,\sigma}$, with $\sigma= \uparrow,\downarrow$, and the one of $\hat\Pi_{th} = \left(\mathds{1}-\sum_{\sigma= \uparrow,\downarrow}\hat\Pi_{\sigma}\right)/(2^L-2)$, the form obtained for $\hat H_{2,\sigma} =4\sum_j \gamma_j \hat P_{j,-\sigma}$, where $-\sigma$ denotes the opposite of the choice made for the value of $\sigma$ and $\hat P_{j,\sigma}$ is the projector onto the state $\ket{\sigma}$ on site $j$, implies that $\hat\Pi_{th}\hat H_2 \hat\Pi_{th}\geq 4 (\min_j \gamma_j) $, thus resulting in a suppression of such coherences at least exponentially fast. The decay of off-diagonal coherences between the sectors identified by the two OSQMBS is in principle even more dramatic, as in such case Eq.~\eqref{Eq:coh_off} must be applied with the choices $\hat\Pi_s = \hat\Pi_{\uparrow}$ and $\hat\Pi_{th} = \hat\Pi_{\downarrow}$, thus resulting in $\hat\Pi_{\downarrow}\hat H_2\hat\Pi_{\downarrow} =4\sum_j \gamma_j$, which in general scales extensively as the system size $L$; thus, the coherences between the sectors spanned by $\ket{F,\uparrow}$ and $\ket{F,\downarrow}$ decay at least exponentially fast with an extensively-scaling decay rate.  

\subsection{Example of tower of OSQMBS} \label{Subsec:tower}

In order to characterize an example of tower of OSQMBS, we borrow the following pair of bond and commutant algebras from Ref.~\cite{Moudgalya_2024_scars}:
\begin{equation}\label{Eq:comm_tower}
    \begin{split}
        \mathcal{A} = &\langle\langle\{\hat S_j^x \hat S_{j+1}^x+\hat S_j^y \hat S_{j+1}^y\},\{ (\hat S_j^z)^2 \},\\
        &\{(\hat S_j^z + \hat S_{j+1}^z)(1-\hat S_j^z \hat S_{j+1}^z)\}, \hat S_{tot}^z  \rangle\rangle    \\
         \mathcal{C}= & \langle\langle\{\ket{\psi_n} \bra{\psi_n} \}_{n=0}^L,\hat S_z^{tot} \rangle\rangle
    \end{split}
\end{equation}
where we introduced the states:
\begin{equation}\label{Eq:tower_spin_1}
 \ket{\psi_n} = \frac{1}{n!\sqrt{\binom{L}{n}}}\left[\sum_j e^{i\pi j} (S_j^+)^2/2\right]^n \otimes_{l=1}^L\ket{-}_l,   
\end{equation}
 denoting the eigenstates of the operator $\hat S_j^z$ with eigenvalues $+1$, $0$ and $-1$ as $\ket{+}_j$, $\ket{0}_j$ and $\ket{-}_j$, respectively, and added the total magnetization operator $\hat S_z^{tot}=\sum_{j=1}^L \hat S_j^z$ to the generators of the bond algebra $\mathcal{A}$ in order to prevent off-diagonal coherences of the form $\ket{\psi_m}\bra{\psi_n}$ with $m\neq n$ to be part of the associated commutant algebra $\mathcal{C}$. 

Let us first distinguish the eigenvalues of $\hat S^z_{tot}$ depending on whether the corresponding sector hosts a state $\ket{\psi_n}$ belonging to the tower of states defined in Eq.~\eqref{Eq:tower_spin_1} or not: we define $S$ as the set containing all eigenvalues of the total magnetization satisfying the former instance, namely $S=\{-L+2n \}_{n=0}^L$, and the set $S^{c}$ as its complement within the spectrum of the total magnetization. The two sets $S$ and $S^c$ therefore satisfy $S\cup S^c =\{-L,\dots,L\}$. Then, the Hilbert space decomposes as:
\begin{equation}
  \begin{split}
   &\mathcal{H}= \left[\bigoplus_{M \in S} \mathcal{H}_M\right]\bigoplus \left[\bigoplus_{M \in S^c} \mathcal{H}_M\right], \\
   & \mathcal{H}_M = \begin{cases}
   \left(\mathcal{H}_{\lambda_{th,M}}^{(\mathcal{A})} \otimes \mathcal{H}_{\lambda_{th,M}}^{(\mathcal{C})} \right)\bigoplus \left(\mathcal{H}_{\lambda_{sc,M}}^{(\mathcal{A})} \otimes \mathcal{H}_{\lambda_{sc,M}}^{(\mathcal{C})} \right) , & M \in S \\
   \left(\mathcal{H}_{\lambda_{th,M}}^{(\mathcal{A})} \otimes \mathcal{H}_{\lambda_{th,M}}^{(\mathcal{C})} \right), & M \in S^c,
   \end{cases}
  \end{split}
\end{equation}
where, as usual, $d_{\lambda_{th,M}}=d_{\lambda_{sc,M}}=1\,\,\forall M$ since $\mathcal{C}$ in Eq.~\eqref{Eq:comm_tower} is abelian, while $D_{\lambda_{th,M}}$ equals the dimension of the magnetization sector with total magnetization $M$, which we denote as $D_M$, when $M\in S^c$, and is equal to $D_M-1$ when $M\in S$ instead. Finally, $D_{\lambda_{sc,M}}=1$, as it equals the dimension of the subspace spanned by the eigentate $\ket{\psi_n}$ with magnetization $M$. 
%Thus, the stationary-state manifold will be described by the set of convex combinations of the projectors $\ket{\psi_n}\bra{\psi_n}$ onto the states belonging to the tower $\{\ket{\psi_n} \}_{n=0}^L$, and of the projectors onto the orthogonal complements of $span\{\ket{\psi_n}\}$ within the corresponding fixed magnetization sector, for each value of $n$, with the projectors onto the fixed-magnetization sectors that do not host a state belonging to the tower. In symbols, a generic stationary state will take the form:

The dynamics within a generic magnetization sector $M_{n_0}=-L+2n_0 \in S$ is analogous to the one presented in presence of a unique OSQMBS, as, by virtue of the result in Eq.~\eqref{Eq:coh_off}, the off-diagonal coherences between the linear span of $\ket{\psi_{n_0}}$ and its orthogonal complement in $\mathcal{H}_{M_{n_0}}$ are expected to generically decay at a rate that is determined by the spectrum of $\hat H_2$ in the range of $\hat\Pi_{(span\{\ket{\psi_{n_0}}\})^{\perp}\cap \mathcal{H}_{M_{n_0}}}$, and the most general expression of the stationary state resulting from an initial state $\hat\rho_0$ that lies in $\mathcal{B}_{M_{n_0},M_{n_0}}=span\{\ket{\psi}\bra{\varphi}:\,\ket{\psi},\,\ket{\varphi}\in \mathcal{H}_{M_{n_0}} \}$ is given by:
\begin{equation}
\begin{split}
    &\hat\rho_{ss}=p \ket{\psi_{n_0}}\bra{\psi_{n_0}} + (1-p)\frac{ \hat\Pi_{(span\{\ket{\psi_{n_0}}\})^{\perp}\cap \mathcal{H}_{M_{n_0}}} }{D_{M_{n_0}} -1},
\end{split}
\end{equation}
where $p=Tr\left[\ket{\psi_{n_0}}\bra{\psi_{n_0}}\hat\rho_0\right]$.

\begin{figure}
\centering
\includegraphics[width=0.5\textwidth]{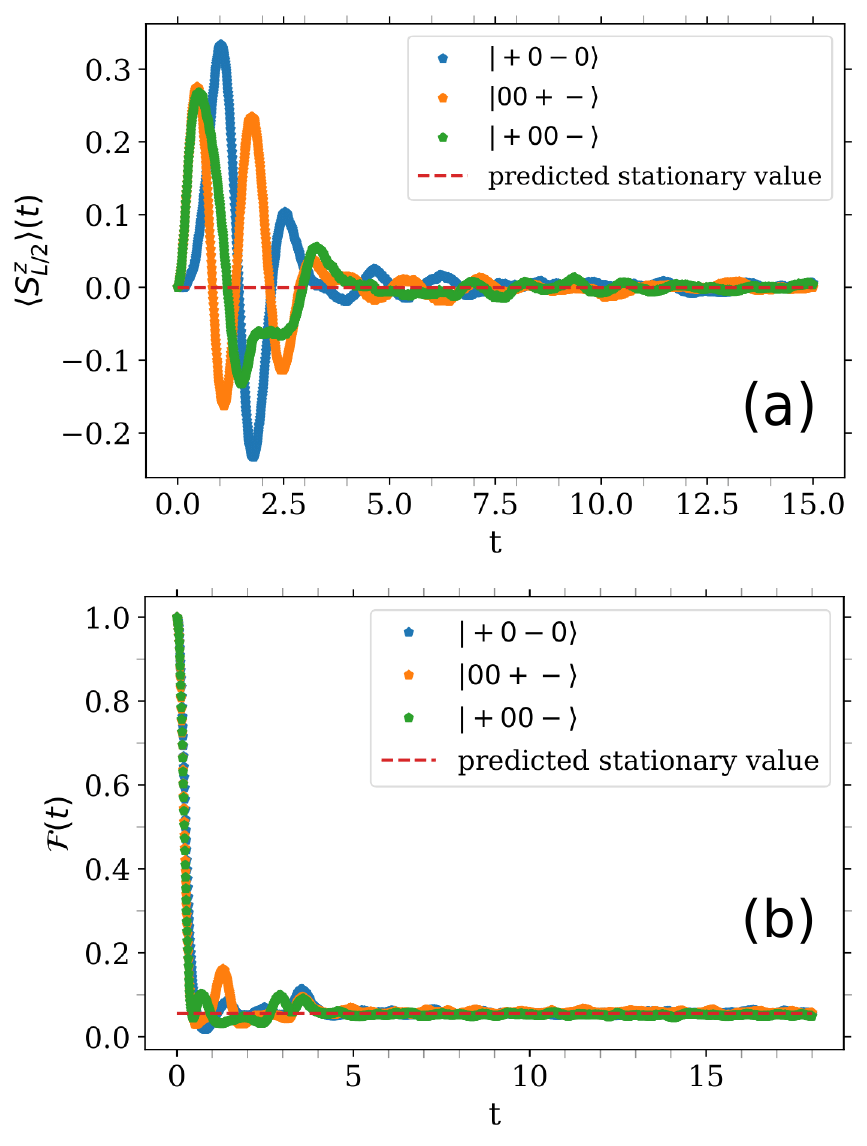}
\caption{(a) Expectation value $\langle \hat S_{L/2}^z \rangle (t) =Tr[\hat S_{L/2}^z \hat\rho(t)]$ as a function of time $t$ for parameter values $(J,D,h,\gamma) = (2.0, 0.2,0.3, 1.0)$ and at size $L=4$. (b) Fidelity $\mathcal{F} (t) =Tr[\ket{\psi(0)}\bra{\psi(0)} \hat\rho(t)]$ as a function of time $t$ for the same parameter values and system size $L$ as in (a).}
\label{Fig:tower}
\end{figure}

Analogously to what done previously, we provide numerical evidence of this phenomenology by considering a specific realization of Lindbladian $\mathcal{L}$ drawn from the algebra $\mathcal{A}$ defined in Eq.~\eqref{Eq:comm_tower} with OBC:
\begin{equation}\label{Eq:lindblad_tower_1}
    \begin{split}
        &\mathcal{L} = -i[\hat H, \cdot]-\frac{\gamma}{2}\sum_{j=1}^L [\hat l_j,[\hat l_j,\cdot]],\\
        & \hat H = J\sum_{j=1}^{L-1}\left(\hat S_j^x \hat S_{j+1}^x +\hat S_j^y \hat S_{j+1}^y \right) +\\
        &+D \sum_{j=1}^{L-1}(\hat S_j^z + \hat S_{j+1}^z)(1-\hat S_j^z \hat S_{j+1}^z)-h \sum_{j=1}^L \hat S_j^z,\\
        & \hat l_j =  \left(\hat S_j^z\right)^2.
    \end{split}
\end{equation}
We estimate the time evolution of the observable $\hat S_{L/2}^z$, namely, $\langle \hat S_{L/2}^z\rangle (t):= Tr[\hat S_{L/2}^z \,\hat \rho(t)]$. From Fig.~\ref{Fig:tower}(a), we conclude that the large-time limit of $\langle \hat S_{L/2}^z\rangle (t)$, when the initial condition lies in the zero-magnetization sector and is orthogonal to the the OSQMBS $\ket{\psi_{L/2}}$, is compatible with the expectation value of the aforesaid observable computed over the typical stationary state in the zero-magnetization sector. The latter reads:
\begin{equation}
\hat\rho_{M=0}:=\frac{\hat\Pi_{\left(span\left\{\ket{\psi_{L/2}}\right\}\right)^{\perp}\cap\, \mathcal{H}_{M=0}}}{D_{M=0}-1 },
\end{equation}
and gives $Tr[\hat S_{L/2}^z\, \hat\rho_{M=0}]=0$, as denoted in Fig.~\ref{Fig:tower}(a). Completely analogous considerations hold for the fidelity $\mathcal{F}(t)=Tr[\ket{\psi(0)}\bra{\psi(0)}\hat\rho(t)]$ presented in Fig.~\ref{Fig:tower}(b), which displays a stationary value that is consistent with the theoretically predicted one, given by $Tr[\ket{\psi(0)}\bra{\psi(0)}\hat\rho_{M=0}] = 1/\left(D_{M=0}-1 \right)$. The latter is found by observing that simple combinatorial considerations allow to derive the result $D_{M=0} = \sum_{k=0}^{\lfloor L/2 \rfloor} \binom{L}{2k}\binom{2k}{k}$. 

Finally, we conclude by showing hints of asymptotic QMBS (AQMBS)~\cite{Gotta_2023} in markovian open quantum systems, which we name asymptotic OSQMBS (AOSQMBS); such exotic states are expected to possess a nontrivial dynamics whose distinguishing feature is the slowdown with increasing system size $L$, in analogy to what observed in closed quantum systems~\cite{Gotta_2023}, where they have been interpreted as slow hydrodynamic modes connected to a nonconventional 'scar symmetry'~\cite{Moudgalya_2024}. Quantitatively, with reference to the tower of states in Eq.~\eqref{Eq:tower_spin_1}, we consider the pure state $\hat\rho_{nk}=\ket{n,k}\bra{n,k}$ corresponding to an AQMBS $\ket{n,k}$, generated by a single application of the creation operator of a momentum-$k$ bimagnon on top of a condensate of $n-1$ momentum-$\pi$ bimagnons, i.e.:
\begin{equation}\label{Eq:aqmbs}
\begin{split}
    &\ket{n,k}=\frac{1}{\sqrt{\mathcal{N}_{n,k}}} \hat J^+_k\ket{\psi_{n-1}},
\end{split}
\end{equation}
where $\hat J^+_k= \frac{1}{2}\sum_j e^{ikj} (\hat S^+_j)^2$, $\mathcal{N}_{n,k}$ is a normalization constant and we fix $k=\pi + 2\pi /L$, which is the choice for which $\ket{n,k}$ exhibits slow dynamics in closed quantum systems.
As a first observation, we characterize the short-time behavior of the expectation value $\langle\hat O_j\rangle(t) = Tr[\hat O_j e^{t\mathcal{L}}(\hat\rho_{nk})]$ of a generic local observable $\hat O_j$ for a time evolution from the initial state $\hat\rho_{nk}$ under a Lindbladian $\mathcal{L}$ constructed from the algebra $\mathcal{A}$ in Eq.~\eqref{Eq:comm_tower}; for the sake of precision, since only the action of the first set of generators of $\mathcal{A}$, namely the exchange terms, on the states defined in Eq.\eqref{Eq:aqmbs} is nontrivial, we assume that such terms appear generically both in the unitary and in the dissipative part of the Lindbladian. We are able to prove the following relation (see SM):
\begin{equation}\label{Eq:time_ev}
    \begin{split}
        &\lim_{L\rightarrow +\infty} \Bigl |\frac{d}{dt}\langle \hat O_j\rangle (t) \Big|_{t=0}\Bigr| = \lim_{L\rightarrow +\infty} \Bigl |Tr[\hat O_j \mathcal{L}(\hat\rho_{nk})] \Bigr|= 0.
    \end{split}
\end{equation}
As a result, one can interpret the vanishing of the first-order corrections to the time-evolution of a generic observable as a distinct signature of the freezing of the dynamics of the state $\hat\rho_{nk}$ in the $L\rightarrow +\infty$ limit. 

\begin{figure}
\centering
\includegraphics[width=0.5\textwidth]{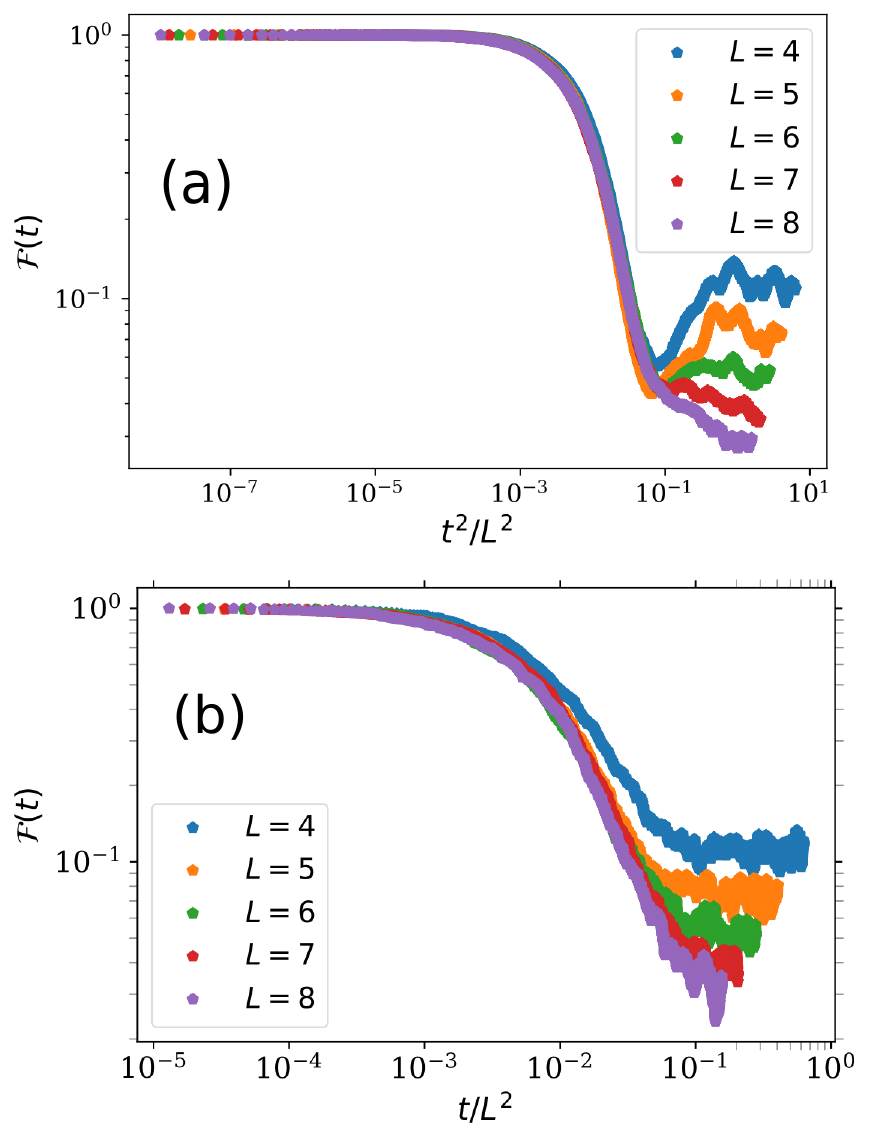}
\caption{(a) Fidelity $\mathcal{F} (t) =Tr[\ket{\psi(0)}\bra{\psi(0)} \hat\rho(t)]$ as a function of $t^2/L^2$ for $\ket{\psi(0)}=\ket{n=1,\, k=\pi+2\pi/L}$ and for parameter values $(J,D,h,\gamma) = (2.0, 0.2,0.3, 1.0)$ in the model of Eq.\eqref{Eq:lindblad_tower_1} and at sizes $L=4,5,6,7,8$. (b) Fidelity $\mathcal{F} (t) =Tr[\ket{\psi(0)}\bra{\psi(0)} \hat\rho(t)]$ as a function of $t/L^2$ for the same initial state $\ket{\psi(0)}$ as in (a) and for parameter values $(D,D_2,h,\gamma) = ( 0.2, 0.8, 1.0, 4.0)$ in the model of Eq.\eqref{Eq:lindblad_tower_2} and at sizes $L=4,5,6,7,8$.}
\label{Fig:fidelities}
\end{figure}

In order to strengthen the analytical observation made in Eq.~\eqref{Eq:time_ev}, we simulate the time evolution of the fidelity $\mathcal{F}(t)=Tr[\hat\rho_{nk} e^{\mathcal{L}t}(\hat\rho_{nk})]$ both with the Lindbladian in Eq.~\eqref{Eq:lindblad_tower_1} and with the following Lindbladian:
\begin{equation}\label{Eq:lindblad_tower_2}
    \begin{split}
        &\mathcal{L} = -i[\hat H, \cdot]-\frac{\gamma}{2}\sum_{j=1}^{L-1} [\hat l_j,[\hat l_j,\cdot]],\\
         \hat H &= D_2\sum_{j=1}^{L}\left(\hat S_j^z\right)^2 +D \sum_{j=1}^{L-1}(\hat S_j^z + \hat S_{j+1}^z)(1-\hat S_j^z \hat S_{j+1}^z)+\\
        &-h \sum_{j=1}^L \hat S_j^z,\\
        & \hat l_j =  \hat S_j^x \hat S_{j+1}^x +\hat S_j^y \hat S_{j+1}^y.
    \end{split}
\end{equation}
A short-time expansion of $\mathcal{F}(t)$ for both Eq.~\eqref{Eq:lindblad_tower_1} and Eq.~\eqref{Eq:lindblad_tower_2}, performed in the SM, suggests a possibly non-perturbative scaling Ansatz of the fidelity to model the anomalously-slow dynamics of the states $\hat \rho_{n,k}$ under the aforesaid Lindbladians within a prethermal time window that follows the start of the quench: while in the case of the model in Eq.~\eqref{Eq:lindblad_tower_1} the Ansatz entails that the fidelity can be described by a universal function $f$ of the ratio $t^2/L^2$ as $\mathcal{F}(t) = f(t^2/L^2)$, in the case of the model in Eq.~\eqref{Eq:lindblad_tower_2} a similar Ansatz in terms of a function $g$ the ratio $t/L^2$ can be proposed through the relation $\mathcal{F}(t) = g(t/L^2)$. We benchmark the proposed scalings by plotting in Fig.~\ref{Fig:fidelities} the fidelity of the state $\hat\rho_{n=1,k}$ as a function of the appropriate space-time ratio for different values of $L$ in the two cases, obtaining an excellent collapse of the different curves in the transient prethermal regime even at the small values of $L$ under numerical investigation and thus confirming the slowdown of the dynamical evolution of OSAQMBS as $L$ increases.

%Its expression for $t\ll 1$ can be manipulated as:
%\begin{equation}
%\begin{split}
%    &|Tr[\rho_{nk}(e^{t\mathcal{L}}\rho_{nk})]|^2 \approx |1+tTr[\rho_{nk}\mathcal{L}\rho_{nk})]|^2 =\\
%    &=|1-t\sum_j\left(\bra{n,k}\hat l^{\dag}_j \hat l_j\ket{n,k}-\bra{n,k}\hat l^{\dag}_j\ket{n,k}\bra{n,k}\hat l_j\ket{n,k}\right)|^2=\\
%    &=|1-t\sum_j ||(\hat l_j-\bra{n,k}\hat l_j\ket{n,k})\ket{n,k}||^2|^2.
%\end{split}
%\end{equation}
%Thus, the short-time decay of the fidelity is controlled by the sum of the variances of the local jump operators on the candidate OAQMBS. We expect this to scale as an inverse power of $L$ in the case of $\hat l_j$ being an operator that annihilates the given tower of states and $\ket{n,k}$ corresponding to a closed-system AQMBS. Moreover, observing that $F'(t)|_{t=0}=2 Re\{ (\rho_{nk},\mathcal{L}\rho_{nk})\}$, we note that assuming the short-time fidelity decay to vanish is equivalent to state that:
%\begin{equation}
%   \lim_{L\rightarrow +\infty}F'(t)|_{t=0} = 0. 
%\end{equation}

Concerning the off-diagonal coherences between the subspace spanned by a scar state $\ket{\psi_{n_0}}$ and its orthogonal complement within the corresponding magnetization sector $M_{n_0}$, which we denote as $W_{n_0}:= M_{n_0}\cap (span\{\ket{\psi_{n_0}}\})^{\perp}$, we leverage the existence of AOSQMBS to showcase an example of slow decay of such coherences that is incompatible with an exponential decay bound with finite decay rate in the thermodynamic limit. Indeed, one may choose the initial pure state:
\begin{equation}\label{Eq:initial_state}
  \begin{split}
      &\hat\rho_0= \ket{\psi_0}\bra{\psi_0},\\
      &\ket{\psi_0} = \frac{1}{\sqrt{2}}\left( \ket{\psi_{n_0}}+\ket{n_0,k}\right).
  \end{split}  
\end{equation}
%and, after expanding the term under derivative sign on the l.h.s. of Eq.~\eqref{Eq:coh_off} to first-order in $t$ in the case $\hat\Pi_s = \ket{\psi_{n_0}}\bra{\psi_{n_0}}$ and $\hat\Pi_{th}= \hat\Pi_{W_{n_0}}$, one obtains:
Since the dissipative term of a Lindbladian is responsible for the loss of off-diagonal coherence in the density matrix, and since the only set of local terms from the bond algebra $\mathcal{A}$ with nonvanishing adjoint action on the state $\ket{n,k}$ are the ones of the form $\{\hat S_j^x \hat S_{j+1}^x + \hat S_j^y \hat S_{j+1}^y \}$, we suppose, for the sake of analytical tractability, that the time evolution of the system is ruled by a Lindbladian $\mathcal{L}$ featuring the latter only among the set of jump operators, while we leave the treatment of the most general case to the SM. Working out the expression of the l.h.s. of Eq.~\eqref{Eq:coh_off} in the present case, one obtains:
\begin{equation}\label{Eq:coh_off_tower}
\begin{split}
    &||\hat\Pi_{W_{n_0}}e^{\mathcal{L}t}(\hat\rho_0)\ket{\psi_{n_0}}\bra{\psi_{n_0}}||^2=\\
    &=\frac{1}{4} \bra{n_0,k}e^{-t \hat H_2}\ket{n_0,k},
\end{split}
\end{equation}
where, in the current scenario, $\hat H_2= \sum_j \gamma_j \left(\hat S_j^x \hat S_{j+1}^x + \hat S_j^y \hat S_{j+1}^y\right)^2$. While a short-time expansion is exploited in the SM to show that this expression is generically inconsistent with an exponential decay of the aforesaid measure of the magnitude of off-diagonal coherences, an exact calculation proving this point can be performed by assuming uniform decay rates ($\gamma_j=\gamma$ $\forall j$) and periodic boundary conditions (PBC) with even values of $L$. In such case, the Hamiltonian $\hat H_2$, when restricted to the effective spin-$1/2$ subspace $\mathcal{S}_{1/2}:=\otimes_{j=1}^L\, span \{\ket{-}_j, \ket{+}_j\}$, takes the form:
\begin{equation}
    \hat H_2\Bigl|_{\mathcal{S}_{1/2}} = 2\gamma \sum_j \left(\frac{1}{4}+\hat\tau_j^x \hat\tau_{j+1}^x +\hat\tau_j^y \hat\tau_{j+1}^y-\hat\tau_j^z \hat\tau_{j+1}^z\right),
\end{equation} 
where $\hat\tau_{j}^{\alpha}$, $\alpha=x,y,z$, are the effective spin-$1/2$ operators within $\mathcal{S}_{1/2}$; it is therefore unitarily equivalent to the Heisenberg Hamiltonian  via a rotation of angle $\pi$ around the z axis of the spins residing on the odd-site sublattice, i.e., $\hat U^{\dag}\,\,\hat H_2\Bigl |_{\mathcal{S}_{1/2}} \hat U= 2\gamma \hat H_{Heis}$, with $\hat U = \prod_{j=0}^{L/2-1} e^{-i\pi \hat \tau_{2j+1}^z}$ and $H_{Heis} = \sum_j\left(\frac{1}{4}-\hat{\vec{\tau}}_j \cdot \hat{\vec{\tau}}_{j+1}\right)$. At the same time, $\hat U^{\dag}\ket{n_0,k}$ is an exact one-magnon eigenstate of $\hat H_{Heis}$, the magnon excitation having momentum $k-\pi =2\pi/L$. All in all, the final result for the time evolution of the expression in Eq.\eqref{Eq:coh_off_tower} takes the form:
\begin{equation}
    ||\hat\Pi_{W_{n_0}}e^{\mathcal{L}t}(\hat\rho_0)\ket{\psi_{n_0}}\bra{\psi_{n_0}}||^2 = \frac{1}{4} \exp(-2\gamma t\epsilon_{k-\pi}),
\end{equation}
where $\epsilon_q = 2\sin^2 (q/2) $ is the one-magnon excitation energy of the Heisenberg model, which scales as $L^{-2}$ for the aforementioned choice $q=k - \pi = 2\pi/L$.

%Since $(\rho_{nk},\mathcal{L}\rho_{nk})= (\rho_{nk},\mathcal{L}_d(\rho_{nk}))= -\sum_m E_m |c_m|^2$, where we have expanded $\rho=\sum_m c_m \rho_m$ in the basis of eigenstates of the dissipative part of the Lindbladian $\mathcal{L}_d=-\frac{1}{2}\sum_j [\hat l_j,[\hat l_j,\cdot]$ with eigenenergy $-E_m$, one can write:
%\begin{equation}
%    (\rho_{nk},\mathcal{L}\rho_{nk})\leq -\min_m E_m \leq 0.
%\end{equation}
%Assuming a well-defined limit for the minimum energy eigenvalue contributing to the above term, one obtains that the dissipative part of the Lindblad superoperator is gapless. We show this explicitly in the appendix in the case of the case of the spin-$1$ XY model.
\section{Conclusions and outlook}

In this work, we have presented an application of the algebraic formalism of commutant algebras to the broad field of the characterization of strong symmetries in markovian open quantum systems, governed by a Lindblad equation. After reviewing how the stationary-state manifold naturally emerges from the analysis of the Hilbert space representations induced by a local algebra and its commutant and applying such observations to recover standard results in spin-chain models with standard discrete symmetries and extensive $U(1)$ symmetries, we have proceeded to a systematic application of the commutant-algebra framework to formulate a formal theory of quantum many-body scarring in markovian open quantum systems. 

More in detail, we have shown the structure that the commutant of a bond algebra must possess for the corresponding class of open systems to display exceptional stationary states, which we dub OSQMBS; we provided numerical signatures of the theoretically predicted properties of the convergence to stationarity and, drawing on symmetry properties, we were able to describe the fate of coherences among the Krylov subspaces identified through the algebra representation. Finally, we gave account of the dynamical freezing phenomena that occur as a result of the existence of OSAQMBS, and their manifestations in the fidelity, observable and off-diagonal coherence dynamics.

Overall, such work sheds light into the very active field of research aimed at the understanding of the impact of symmetries on quantum many-body dynamics, both in closed and open systems. Concerning direct extensions of the results presented in the present work, the set of questions raised by such a work revolve mainly around the results in Section~\ref{Sec:4}, where the commutant-algebra framework showcases its usefulness in understanding the role of unconventional nonlocal strong symmetries in open quantum system dynamics; it would be interesting to further describe, e.g., whether an interface dynamics in analogy to the one studied in Ref.\cite{Marche_2024} can be satisfactorily described in presence of a tower of OSQMBS, and which universality class the space-time propagation of such interface belongs to.

Moreover, it would be instructive to fully explore the consequences of the random-circuit framework introduced in Sec.~\ref{Subsec:brownian}, to investigate hydrodynamic modes and their impact on late-time relaxation of observables, purity and mixedness properties of open quantum many-body systems, in analogy to the corresponding work carried out in the realm of closed quantum systems~\cite{Moudgalya_2024,Vardhan_2024}.

\paragraph*{Acknowledgements ---}
 I deeply thank Leonardo Mazza for the several suggestions and comments on the manuscript and Yahui Li for a fruitful discussion. I also am in debt to Alice Marché, Gianluca Morettini, Luca Capizzi, Leonardo Mazza and Sanjay Moudgalya for previous collaborations on strictly related subjects. I thank Thierry Giamarchi for his support.
 This work was supported by the Swiss National Science Foundation under Division II
(Grant No. 200020-219400).

\bibliography{osqs_bib}

\begin{thebibliography}{37}%
\makeatletter
\providecommand \@ifxundefined [1]{%
 \@ifx{#1\undefined}
}%
\providecommand \@ifnum [1]{%
 \ifnum #1\expandafter \@firstoftwo
 \else \expandafter \@secondoftwo
 \fi
}%
\providecommand \@ifx [1]{%
 \ifx #1\expandafter \@firstoftwo
 \else \expandafter \@secondoftwo
 \fi
}%
\providecommand \natexlab [1]{#1}%
\providecommand \enquote  [1]{``#1''}%
\providecommand \bibnamefont  [1]{#1}%
\providecommand \bibfnamefont [1]{#1}%
\providecommand \citenamefont [1]{#1}%
\providecommand \href@noop [0]{\@secondoftwo}%
\providecommand \href [0]{\begingroup \@sanitize@url \@href}%
\providecommand \@href[1]{\@@startlink{#1}\@@href}%
\providecommand \@@href[1]{\endgroup#1\@@endlink}%
\providecommand \@sanitize@url [0]{\catcode `\\12\catcode `\$12\catcode
  `\&12\catcode `\#12\catcode `\^12\catcode `\_12\catcode `\%12\relax}%
\providecommand \@@startlink[1]{}%
\providecommand \@@endlink[0]{}%
\providecommand \url  [0]{\begingroup\@sanitize@url \@url }%
\providecommand \@url [1]{\endgroup\@href {#1}{\urlprefix }}%
\providecommand \urlprefix  [0]{URL }%
\providecommand \Eprint [0]{\href }%
\providecommand \doibase [0]{https://doi.org/}%
\providecommand \selectlanguage [0]{\@gobble}%
\providecommand \bibinfo  [0]{\@secondoftwo}%
\providecommand \bibfield  [0]{\@secondoftwo}%
\providecommand \translation [1]{[#1]}%
\providecommand \BibitemOpen [0]{}%
\providecommand \bibitemStop [0]{}%
\providecommand \bibitemNoStop [0]{.\EOS\space}%
\providecommand \EOS [0]{\spacefactor3000\relax}%
\providecommand \BibitemShut  [1]{\csname bibitem#1\endcsname}%
\let\auto@bib@innerbib\@empty
%</preamble>
\bibitem [{\citenamefont {Ogunnaike}\ \emph {et~al.}(2023)\citenamefont
  {Ogunnaike}, \citenamefont {Feldmeier},\ and\ \citenamefont
  {Lee}}]{Ogunnaike_2023}%
  \BibitemOpen
  \bibfield  {author} {\bibinfo {author} {\bibfnamefont {O.}~\bibnamefont
  {Ogunnaike}}, \bibinfo {author} {\bibfnamefont {J.}~\bibnamefont
  {Feldmeier}},\ and\ \bibinfo {author} {\bibfnamefont {J.~Y.}\ \bibnamefont
  {Lee}},\ }\bibfield  {title} {\bibinfo {title} {Unifying emergent
  hydrodynamics and lindbladian low-energy spectra across symmetries,
  constraints, and long-range interactions},\ }\href
  {https://doi.org/10.1103/PhysRevLett.131.220403} {\bibfield  {journal}
  {\bibinfo  {journal} {Phys. Rev. Lett.}\ }\textbf {\bibinfo {volume} {131}},\
  \bibinfo {pages} {220403} (\bibinfo {year} {2023})}\BibitemShut {NoStop}%
\bibitem [{\citenamefont {Moudgalya}\ and\ \citenamefont
  {Motrunich}(2024{\natexlab{a}})}]{Moudgalya_2024}%
  \BibitemOpen
  \bibfield  {author} {\bibinfo {author} {\bibfnamefont {S.}~\bibnamefont
  {Moudgalya}}\ and\ \bibinfo {author} {\bibfnamefont {O.~I.}\ \bibnamefont
  {Motrunich}},\ }\bibfield  {title} {\bibinfo {title} {Symmetries as ground
  states of local superoperators: Hydrodynamic implications},\ }\href
  {https://doi.org/10.1103/PRXQuantum.5.040330} {\bibfield  {journal} {\bibinfo
   {journal} {PRX Quantum}\ }\textbf {\bibinfo {volume} {5}},\ \bibinfo {pages}
  {040330} (\bibinfo {year} {2024}{\natexlab{a}})}\BibitemShut {NoStop}%
\bibitem [{\citenamefont {Rigol}\ \emph {et~al.}(2008)\citenamefont {Rigol},
  \citenamefont {Dunjko},\ and\ \citenamefont {Olshanii}}]{Rigol_2008}%
  \BibitemOpen
  \bibfield  {author} {\bibinfo {author} {\bibfnamefont {M.}~\bibnamefont
  {Rigol}}, \bibinfo {author} {\bibfnamefont {V.}~\bibnamefont {Dunjko}},\ and\
  \bibinfo {author} {\bibfnamefont {M.}~\bibnamefont {Olshanii}},\ }\bibfield
  {title} {\bibinfo {title} {Thermalization and its mechanism for generic
  isolated quantum systems},\ }\href@noop {} {\bibfield  {journal} {\bibinfo
  {journal} {Nature}\ }\textbf {\bibinfo {volume} {452}},\ \bibinfo {pages}
  {854} (\bibinfo {year} {2008})}\BibitemShut {NoStop}%
\bibitem [{\citenamefont {Daley}(2014)}]{Daley_2014}%
  \BibitemOpen
  \bibfield  {author} {\bibinfo {author} {\bibfnamefont {A.~J.}\ \bibnamefont
  {Daley}},\ }\bibfield  {title} {\bibinfo {title} {Quantum trajectories and
  open many-body quantum systems},\ }\href@noop {} {\bibfield  {journal}
  {\bibinfo  {journal} {Advances in Physics}\ }\textbf {\bibinfo {volume}
  {63}},\ \bibinfo {pages} {77} (\bibinfo {year} {2014})}\BibitemShut {NoStop}%
\bibitem [{\citenamefont {Fazio}\ \emph {et~al.}(2025)\citenamefont {Fazio},
  \citenamefont {Keeling}, \citenamefont {Mazza},\ and\ \citenamefont
  {Schir{\`o}}}]{Fazio_2025}%
  \BibitemOpen
  \bibfield  {author} {\bibinfo {author} {\bibfnamefont {R.}~\bibnamefont
  {Fazio}}, \bibinfo {author} {\bibfnamefont {J.}~\bibnamefont {Keeling}},
  \bibinfo {author} {\bibfnamefont {L.}~\bibnamefont {Mazza}},\ and\ \bibinfo
  {author} {\bibfnamefont {M.}~\bibnamefont {Schir{\`o}}},\ }\bibfield  {title}
  {\bibinfo {title} {Many-body open quantum systems},\ }\href@noop {}
  {\bibfield  {journal} {\bibinfo  {journal} {SciPost Physics Lecture Notes}\
  ,\ \bibinfo {pages} {099}} (\bibinfo {year} {2025})}\BibitemShut {NoStop}%
\bibitem [{\citenamefont {Bu{\v{c}}a}\ and\ \citenamefont
  {Prosen}(2012)}]{Buca_2012}%
  \BibitemOpen
  \bibfield  {author} {\bibinfo {author} {\bibfnamefont {B.}~\bibnamefont
  {Bu{\v{c}}a}}\ and\ \bibinfo {author} {\bibfnamefont {T.}~\bibnamefont
  {Prosen}},\ }\bibfield  {title} {\bibinfo {title} {A note on symmetry
  reductions of the lindblad equation: transport in constrained open spin
  chains},\ }\href@noop {} {\bibfield  {journal} {\bibinfo  {journal} {New
  Journal of Physics}\ }\textbf {\bibinfo {volume} {14}},\ \bibinfo {pages}
  {073007} (\bibinfo {year} {2012})}\BibitemShut {NoStop}%
\bibitem [{\citenamefont {Albert}\ and\ \citenamefont
  {Jiang}(2014)}]{Albert_2014}%
  \BibitemOpen
  \bibfield  {author} {\bibinfo {author} {\bibfnamefont {V.~V.}\ \bibnamefont
  {Albert}}\ and\ \bibinfo {author} {\bibfnamefont {L.}~\bibnamefont {Jiang}},\
  }\bibfield  {title} {\bibinfo {title} {Symmetries and conserved quantities in
  lindblad master equations},\ }\href
  {https://doi.org/10.1103/PhysRevA.89.022118} {\bibfield  {journal} {\bibinfo
  {journal} {Phys. Rev. A}\ }\textbf {\bibinfo {volume} {89}},\ \bibinfo
  {pages} {022118} (\bibinfo {year} {2014})}\BibitemShut {NoStop}%
\bibitem [{\citenamefont {Zhang}\ \emph {et~al.}(2020)\citenamefont {Zhang},
  \citenamefont {Tindall}, \citenamefont {Mur-Petit}, \citenamefont {Jaksch},\
  and\ \citenamefont {Bu{\v{c}}a}}]{Zhang_2020}%
  \BibitemOpen
  \bibfield  {author} {\bibinfo {author} {\bibfnamefont {Z.}~\bibnamefont
  {Zhang}}, \bibinfo {author} {\bibfnamefont {J.}~\bibnamefont {Tindall}},
  \bibinfo {author} {\bibfnamefont {J.}~\bibnamefont {Mur-Petit}}, \bibinfo
  {author} {\bibfnamefont {D.}~\bibnamefont {Jaksch}},\ and\ \bibinfo {author}
  {\bibfnamefont {B.}~\bibnamefont {Bu{\v{c}}a}},\ }\bibfield  {title}
  {\bibinfo {title} {Stationary state degeneracy of open quantum systems with
  non-abelian symmetries},\ }\href@noop {} {\bibfield  {journal} {\bibinfo
  {journal} {Journal of Physics A: Mathematical and Theoretical}\ }\textbf
  {\bibinfo {volume} {53}},\ \bibinfo {pages} {215304} (\bibinfo {year}
  {2020})}\BibitemShut {NoStop}%
\bibitem [{\citenamefont {Nigro}(2019)}]{Nigro_2019}%
  \BibitemOpen
  \bibfield  {author} {\bibinfo {author} {\bibfnamefont {D.}~\bibnamefont
  {Nigro}},\ }\bibfield  {title} {\bibinfo {title} {On the uniqueness of the
  steady-state solution of the lindblad--gorini--kossakowski--sudarshan
  equation},\ }\href@noop {} {\bibfield  {journal} {\bibinfo  {journal}
  {Journal of Statistical Mechanics: Theory and Experiment}\ }\textbf {\bibinfo
  {volume} {2019}},\ \bibinfo {pages} {043202} (\bibinfo {year}
  {2019})}\BibitemShut {NoStop}%
\bibitem [{\citenamefont {Yoshida}(2024)}]{Yoshida_2024}%
  \BibitemOpen
  \bibfield  {author} {\bibinfo {author} {\bibfnamefont {H.}~\bibnamefont
  {Yoshida}},\ }\bibfield  {title} {\bibinfo {title} {Uniqueness of steady
  states of gorini-kossakowski-sudarshan-lindblad equations: A simple proof},\
  }\href {https://doi.org/10.1103/PhysRevA.109.022218} {\bibfield  {journal}
  {\bibinfo  {journal} {Phys. Rev. A}\ }\textbf {\bibinfo {volume} {109}},\
  \bibinfo {pages} {022218} (\bibinfo {year} {2024})}\BibitemShut {NoStop}%
\bibitem [{\citenamefont {Moudgalya}\ and\ \citenamefont
  {Motrunich}(2022)}]{Moudgalya_2022}%
  \BibitemOpen
  \bibfield  {author} {\bibinfo {author} {\bibfnamefont {S.}~\bibnamefont
  {Moudgalya}}\ and\ \bibinfo {author} {\bibfnamefont {O.~I.}\ \bibnamefont
  {Motrunich}},\ }\bibfield  {title} {\bibinfo {title} {Hilbert space
  fragmentation and commutant algebras},\ }\href
  {https://doi.org/10.1103/PhysRevX.12.011050} {\bibfield  {journal} {\bibinfo
  {journal} {Phys. Rev. X}\ }\textbf {\bibinfo {volume} {12}},\ \bibinfo
  {pages} {011050} (\bibinfo {year} {2022})}\BibitemShut {NoStop}%
\bibitem [{\citenamefont {Moudgalya}\ and\ \citenamefont
  {Motrunich}(2023)}]{Moudgalya_2023}%
  \BibitemOpen
  \bibfield  {author} {\bibinfo {author} {\bibfnamefont {S.}~\bibnamefont
  {Moudgalya}}\ and\ \bibinfo {author} {\bibfnamefont {O.~I.}\ \bibnamefont
  {Motrunich}},\ }\bibfield  {title} {\bibinfo {title} {From symmetries to
  commutant algebras in standard hamiltonians},\ }\href
  {https://doi.org/https://doi.org/10.1016/j.aop.2023.169384} {\bibfield
  {journal} {\bibinfo  {journal} {Annals of Physics}\ }\textbf {\bibinfo
  {volume} {455}},\ \bibinfo {pages} {169384} (\bibinfo {year}
  {2023})}\BibitemShut {NoStop}%
\bibitem [{\citenamefont {Moudgalya}\ and\ \citenamefont
  {Motrunich}(2024{\natexlab{b}})}]{Moudgalya_2024_scars}%
  \BibitemOpen
  \bibfield  {author} {\bibinfo {author} {\bibfnamefont {S.}~\bibnamefont
  {Moudgalya}}\ and\ \bibinfo {author} {\bibfnamefont {O.~I.}\ \bibnamefont
  {Motrunich}},\ }\bibfield  {title} {\bibinfo {title} {Exhaustive
  characterization of quantum many-body scars using commutant algebras},\
  }\href {https://doi.org/10.1103/PhysRevX.14.041069} {\bibfield  {journal}
  {\bibinfo  {journal} {Phys. Rev. X}\ }\textbf {\bibinfo {volume} {14}},\
  \bibinfo {pages} {041069} (\bibinfo {year} {2024}{\natexlab{b}})}\BibitemShut
  {NoStop}%
\bibitem [{\citenamefont {Li}\ \emph {et~al.}(2023)\citenamefont {Li},
  \citenamefont {Sala},\ and\ \citenamefont {Pollmann}}]{Li_2023}%
  \BibitemOpen
  \bibfield  {author} {\bibinfo {author} {\bibfnamefont {Y.}~\bibnamefont
  {Li}}, \bibinfo {author} {\bibfnamefont {P.}~\bibnamefont {Sala}},\ and\
  \bibinfo {author} {\bibfnamefont {F.}~\bibnamefont {Pollmann}},\ }\bibfield
  {title} {\bibinfo {title} {Hilbert space fragmentation in open quantum
  systems},\ }\href {https://doi.org/10.1103/PhysRevResearch.5.043239}
  {\bibfield  {journal} {\bibinfo  {journal} {Phys. Rev. Res.}\ }\textbf
  {\bibinfo {volume} {5}},\ \bibinfo {pages} {043239} (\bibinfo {year}
  {2023})}\BibitemShut {NoStop}%
\bibitem [{\citenamefont {Li}\ \emph {et~al.}(2025)\citenamefont {Li},
  \citenamefont {Pollmann}, \citenamefont {Read},\ and\ \citenamefont
  {Sala}}]{Li_2025}%
  \BibitemOpen
  \bibfield  {author} {\bibinfo {author} {\bibfnamefont {Y.}~\bibnamefont
  {Li}}, \bibinfo {author} {\bibfnamefont {F.}~\bibnamefont {Pollmann}},
  \bibinfo {author} {\bibfnamefont {N.}~\bibnamefont {Read}},\ and\ \bibinfo
  {author} {\bibfnamefont {P.}~\bibnamefont {Sala}},\ }\bibfield  {title}
  {\bibinfo {title} {Highly entangled stationary states from strong
  symmetries},\ }\href {https://doi.org/10.1103/PhysRevX.15.011068} {\bibfield
  {journal} {\bibinfo  {journal} {Phys. Rev. X}\ }\textbf {\bibinfo {volume}
  {15}},\ \bibinfo {pages} {011068} (\bibinfo {year} {2025})}\BibitemShut
  {NoStop}%
\bibitem [{\citenamefont {Paszko}\ \emph {et~al.}(2025)\citenamefont {Paszko},
  \citenamefont {Turner}, \citenamefont {Rose},\ and\ \citenamefont
  {Pal}}]{Paszko_2025}%
  \BibitemOpen
  \bibfield  {author} {\bibinfo {author} {\bibfnamefont {D.}~\bibnamefont
  {Paszko}}, \bibinfo {author} {\bibfnamefont {C.~J.}\ \bibnamefont {Turner}},
  \bibinfo {author} {\bibfnamefont {D.~C.}\ \bibnamefont {Rose}},\ and\
  \bibinfo {author} {\bibfnamefont {A.}~\bibnamefont {Pal}},\ }\bibfield
  {title} {\bibinfo {title} {Operator-space fragmentation and integrability in
  pauli-lindblad models},\ }\href@noop {} {\bibfield  {journal} {\bibinfo
  {journal} {arXiv preprint arXiv:2506.16518}\ } (\bibinfo {year}
  {2025})}\BibitemShut {NoStop}%
\bibitem [{\citenamefont {March\'e}\ \emph {et~al.}(2025)\citenamefont
  {March\'e}, \citenamefont {Morettini}, \citenamefont {Mazza}, \citenamefont
  {Gotta},\ and\ \citenamefont {Capizzi}}]{Marche_2024}%
  \BibitemOpen
  \bibfield  {author} {\bibinfo {author} {\bibfnamefont {A.}~\bibnamefont
  {March\'e}}, \bibinfo {author} {\bibfnamefont {G.}~\bibnamefont {Morettini}},
  \bibinfo {author} {\bibfnamefont {L.}~\bibnamefont {Mazza}}, \bibinfo
  {author} {\bibfnamefont {L.}~\bibnamefont {Gotta}},\ and\ \bibinfo {author}
  {\bibfnamefont {L.}~\bibnamefont {Capizzi}},\ }\bibfield  {title} {\bibinfo
  {title} {Exceptional stationary state in a dephasing many-body open quantum
  system},\ }\href {https://doi.org/10.1103/zn9v-k73w} {\bibfield  {journal}
  {\bibinfo  {journal} {Phys. Rev. Lett.}\ }\textbf {\bibinfo {volume} {135}},\
  \bibinfo {pages} {020406} (\bibinfo {year} {2025})}\BibitemShut {NoStop}%
\bibitem [{\citenamefont {Mazur}(1969)}]{Mazur_1969}%
  \BibitemOpen
  \bibfield  {author} {\bibinfo {author} {\bibfnamefont {P.}~\bibnamefont
  {Mazur}},\ }\bibfield  {title} {\bibinfo {title} {Non-ergodicity of phase
  functions in certain systems},\ }\href@noop {} {\bibfield  {journal}
  {\bibinfo  {journal} {Physica}\ }\textbf {\bibinfo {volume} {43}},\ \bibinfo
  {pages} {533} (\bibinfo {year} {1969})}\BibitemShut {NoStop}%
\bibitem [{\citenamefont {Dhar}\ \emph {et~al.}(2021)\citenamefont {Dhar},
  \citenamefont {Kundu},\ and\ \citenamefont {Saito}}]{Dhar_2021}%
  \BibitemOpen
  \bibfield  {author} {\bibinfo {author} {\bibfnamefont {A.}~\bibnamefont
  {Dhar}}, \bibinfo {author} {\bibfnamefont {A.}~\bibnamefont {Kundu}},\ and\
  \bibinfo {author} {\bibfnamefont {K.}~\bibnamefont {Saito}},\ }\bibfield
  {title} {\bibinfo {title} {Revisiting the mazur bound and the suzuki
  equality},\ }\href@noop {} {\bibfield  {journal} {\bibinfo  {journal} {Chaos,
  Solitons \& Fractals}\ }\textbf {\bibinfo {volume} {144}},\ \bibinfo {pages}
  {110618} (\bibinfo {year} {2021})}\BibitemShut {NoStop}%
\bibitem [{\citenamefont {Turner}\ \emph {et~al.}(2018)\citenamefont {Turner},
  \citenamefont {Michailidis}, \citenamefont {Abanin}, \citenamefont {Serbyn},\
  and\ \citenamefont {Papi\ifmmode~\acute{c}\else \'{c}\fi{}}}]{Turner_2018}%
  \BibitemOpen
  \bibfield  {author} {\bibinfo {author} {\bibfnamefont {C.~J.}\ \bibnamefont
  {Turner}}, \bibinfo {author} {\bibfnamefont {A.~A.}\ \bibnamefont
  {Michailidis}}, \bibinfo {author} {\bibfnamefont {D.~A.}\ \bibnamefont
  {Abanin}}, \bibinfo {author} {\bibfnamefont {M.}~\bibnamefont {Serbyn}},\
  and\ \bibinfo {author} {\bibfnamefont {Z.}~\bibnamefont
  {Papi\ifmmode~\acute{c}\else \'{c}\fi{}}},\ }\bibfield  {title} {\bibinfo
  {title} {Quantum scarred eigenstates in a rydberg atom chain: Entanglement,
  breakdown of thermalization, and stability to perturbations},\ }\href
  {https://doi.org/10.1103/PhysRevB.98.155134} {\bibfield  {journal} {\bibinfo
  {journal} {Phys. Rev. B}\ }\textbf {\bibinfo {volume} {98}},\ \bibinfo
  {pages} {155134} (\bibinfo {year} {2018})}\BibitemShut {NoStop}%
\bibitem [{Tur(2018)}]{Turner_2018_2}%
  \BibitemOpen
  \bibfield  {title} {\bibinfo {title} {Weak ergodicity breaking from quantum
  many-body scars},\ }\href {https://doi.org/10.1038/s41567-018-0137-5}
  {\bibfield  {journal} {\bibinfo  {journal} {Nature Physics}\ }\textbf
  {\bibinfo {volume} {14}},\ \bibinfo {pages} {745} (\bibinfo {year}
  {2018})}\BibitemShut {NoStop}%
\bibitem [{\citenamefont {Choi}\ \emph {et~al.}(2019)\citenamefont {Choi},
  \citenamefont {Turner}, \citenamefont {Pichler}, \citenamefont {Ho},
  \citenamefont {Michailidis}, \citenamefont {Papi\ifmmode~\acute{c}\else
  \'{c}\fi{}}, \citenamefont {Serbyn}, \citenamefont {Lukin},\ and\
  \citenamefont {Abanin}}]{Choi_2019}%
  \BibitemOpen
  \bibfield  {author} {\bibinfo {author} {\bibfnamefont {S.}~\bibnamefont
  {Choi}}, \bibinfo {author} {\bibfnamefont {C.~J.}\ \bibnamefont {Turner}},
  \bibinfo {author} {\bibfnamefont {H.}~\bibnamefont {Pichler}}, \bibinfo
  {author} {\bibfnamefont {W.~W.}\ \bibnamefont {Ho}}, \bibinfo {author}
  {\bibfnamefont {A.~A.}\ \bibnamefont {Michailidis}}, \bibinfo {author}
  {\bibfnamefont {Z.}~\bibnamefont {Papi\ifmmode~\acute{c}\else \'{c}\fi{}}},
  \bibinfo {author} {\bibfnamefont {M.}~\bibnamefont {Serbyn}}, \bibinfo
  {author} {\bibfnamefont {M.~D.}\ \bibnamefont {Lukin}},\ and\ \bibinfo
  {author} {\bibfnamefont {D.~A.}\ \bibnamefont {Abanin}},\ }\bibfield  {title}
  {\bibinfo {title} {Emergent su(2) dynamics and perfect quantum many-body
  scars},\ }\href {https://doi.org/10.1103/PhysRevLett.122.220603} {\bibfield
  {journal} {\bibinfo  {journal} {Phys. Rev. Lett.}\ }\textbf {\bibinfo
  {volume} {122}},\ \bibinfo {pages} {220603} (\bibinfo {year}
  {2019})}\BibitemShut {NoStop}%
\bibitem [{\citenamefont {Shibata}\ \emph {et~al.}(2020)\citenamefont
  {Shibata}, \citenamefont {Yoshioka},\ and\ \citenamefont
  {Katsura}}]{Shibata_2020}%
  \BibitemOpen
  \bibfield  {author} {\bibinfo {author} {\bibfnamefont {N.}~\bibnamefont
  {Shibata}}, \bibinfo {author} {\bibfnamefont {N.}~\bibnamefont {Yoshioka}},\
  and\ \bibinfo {author} {\bibfnamefont {H.}~\bibnamefont {Katsura}},\
  }\bibfield  {title} {\bibinfo {title} {Onsager's scars in disordered spin
  chains},\ }\href {https://doi.org/10.1103/PhysRevLett.124.180604} {\bibfield
  {journal} {\bibinfo  {journal} {Phys. Rev. Lett.}\ }\textbf {\bibinfo
  {volume} {124}},\ \bibinfo {pages} {180604} (\bibinfo {year}
  {2020})}\BibitemShut {NoStop}%
\bibitem [{\citenamefont {Moudgalya}\ \emph
  {et~al.}(2020{\natexlab{a}})\citenamefont {Moudgalya}, \citenamefont
  {Regnault},\ and\ \citenamefont {Bernevig}}]{Moudgalya_2020}%
  \BibitemOpen
  \bibfield  {author} {\bibinfo {author} {\bibfnamefont {S.}~\bibnamefont
  {Moudgalya}}, \bibinfo {author} {\bibfnamefont {N.}~\bibnamefont
  {Regnault}},\ and\ \bibinfo {author} {\bibfnamefont {B.~A.}\ \bibnamefont
  {Bernevig}},\ }\bibfield  {title} {\bibinfo {title}
  {$\ensuremath{\eta}$-pairing in hubbard models: From spectrum generating
  algebras to quantum many-body scars},\ }\href
  {https://doi.org/10.1103/PhysRevB.102.085140} {\bibfield  {journal} {\bibinfo
   {journal} {Phys. Rev. B}\ }\textbf {\bibinfo {volume} {102}},\ \bibinfo
  {pages} {085140} (\bibinfo {year} {2020}{\natexlab{a}})}\BibitemShut
  {NoStop}%
\bibitem [{\citenamefont {Moudgalya}\ \emph
  {et~al.}(2020{\natexlab{b}})\citenamefont {Moudgalya}, \citenamefont
  {O'Brien}, \citenamefont {Bernevig}, \citenamefont {Fendley},\ and\
  \citenamefont {Regnault}}]{Moudgalya_2020_2}%
  \BibitemOpen
  \bibfield  {author} {\bibinfo {author} {\bibfnamefont {S.}~\bibnamefont
  {Moudgalya}}, \bibinfo {author} {\bibfnamefont {E.}~\bibnamefont {O'Brien}},
  \bibinfo {author} {\bibfnamefont {B.~A.}\ \bibnamefont {Bernevig}}, \bibinfo
  {author} {\bibfnamefont {P.}~\bibnamefont {Fendley}},\ and\ \bibinfo {author}
  {\bibfnamefont {N.}~\bibnamefont {Regnault}},\ }\bibfield  {title} {\bibinfo
  {title} {Large classes of quantum scarred hamiltonians from matrix product
  states},\ }\href {https://doi.org/10.1103/PhysRevB.102.085120} {\bibfield
  {journal} {\bibinfo  {journal} {Phys. Rev. B}\ }\textbf {\bibinfo {volume}
  {102}},\ \bibinfo {pages} {085120} (\bibinfo {year}
  {2020}{\natexlab{b}})}\BibitemShut {NoStop}%
\bibitem [{\citenamefont {Moudgalya}\ \emph
  {et~al.}(2020{\natexlab{c}})\citenamefont {Moudgalya}, \citenamefont
  {Bernevig},\ and\ \citenamefont {Regnault}}]{Moudgalya_2020_3}%
  \BibitemOpen
  \bibfield  {author} {\bibinfo {author} {\bibfnamefont {S.}~\bibnamefont
  {Moudgalya}}, \bibinfo {author} {\bibfnamefont {B.~A.}\ \bibnamefont
  {Bernevig}},\ and\ \bibinfo {author} {\bibfnamefont {N.}~\bibnamefont
  {Regnault}},\ }\bibfield  {title} {\bibinfo {title} {Quantum many-body scars
  in a landau level on a thin torus},\ }\href
  {https://doi.org/10.1103/PhysRevB.102.195150} {\bibfield  {journal} {\bibinfo
   {journal} {Phys. Rev. B}\ }\textbf {\bibinfo {volume} {102}},\ \bibinfo
  {pages} {195150} (\bibinfo {year} {2020}{\natexlab{c}})}\BibitemShut
  {NoStop}%
\bibitem [{\citenamefont {Iadecola}\ \emph {et~al.}(2019)\citenamefont
  {Iadecola}, \citenamefont {Schecter},\ and\ \citenamefont
  {Xu}}]{Iadecola_2019}%
  \BibitemOpen
  \bibfield  {author} {\bibinfo {author} {\bibfnamefont {T.}~\bibnamefont
  {Iadecola}}, \bibinfo {author} {\bibfnamefont {M.}~\bibnamefont {Schecter}},\
  and\ \bibinfo {author} {\bibfnamefont {S.}~\bibnamefont {Xu}},\ }\bibfield
  {title} {\bibinfo {title} {Quantum many-body scars from magnon
  condensation},\ }\href {https://doi.org/10.1103/PhysRevB.100.184312}
  {\bibfield  {journal} {\bibinfo  {journal} {Phys. Rev. B}\ }\textbf {\bibinfo
  {volume} {100}},\ \bibinfo {pages} {184312} (\bibinfo {year}
  {2019})}\BibitemShut {NoStop}%
\bibitem [{\citenamefont {Iadecola}\ and\ \citenamefont
  {Schecter}(2020)}]{Iadecola_2020}%
  \BibitemOpen
  \bibfield  {author} {\bibinfo {author} {\bibfnamefont {T.}~\bibnamefont
  {Iadecola}}\ and\ \bibinfo {author} {\bibfnamefont {M.}~\bibnamefont
  {Schecter}},\ }\bibfield  {title} {\bibinfo {title} {Quantum many-body scar
  states with emergent kinetic constraints and finite-entanglement revivals},\
  }\href {https://doi.org/10.1103/PhysRevB.101.024306} {\bibfield  {journal}
  {\bibinfo  {journal} {Phys. Rev. B}\ }\textbf {\bibinfo {volume} {101}},\
  \bibinfo {pages} {024306} (\bibinfo {year} {2020})}\BibitemShut {NoStop}%
\bibitem [{\citenamefont {Schecter}\ and\ \citenamefont
  {Iadecola}(2019)}]{Schecter_2019}%
  \BibitemOpen
  \bibfield  {author} {\bibinfo {author} {\bibfnamefont {M.}~\bibnamefont
  {Schecter}}\ and\ \bibinfo {author} {\bibfnamefont {T.}~\bibnamefont
  {Iadecola}},\ }\bibfield  {title} {\bibinfo {title} {Weak ergodicity breaking
  and quantum many-body scars in spin-1 $xy$ magnets},\ }\href
  {https://doi.org/10.1103/PhysRevLett.123.147201} {\bibfield  {journal}
  {\bibinfo  {journal} {Phys. Rev. Lett.}\ }\textbf {\bibinfo {volume} {123}},\
  \bibinfo {pages} {147201} (\bibinfo {year} {2019})}\BibitemShut {NoStop}%
\bibitem [{\citenamefont {Gotta}\ \emph {et~al.}(2022)\citenamefont {Gotta},
  \citenamefont {Mazza}, \citenamefont {Simon},\ and\ \citenamefont
  {Roux}}]{Gotta_2022}%
  \BibitemOpen
  \bibfield  {author} {\bibinfo {author} {\bibfnamefont {L.}~\bibnamefont
  {Gotta}}, \bibinfo {author} {\bibfnamefont {L.}~\bibnamefont {Mazza}},
  \bibinfo {author} {\bibfnamefont {P.}~\bibnamefont {Simon}},\ and\ \bibinfo
  {author} {\bibfnamefont {G.}~\bibnamefont {Roux}},\ }\bibfield  {title}
  {\bibinfo {title} {Exact many-body scars based on pairs or multimers in a
  chain of spinless fermions},\ }\href
  {https://doi.org/10.1103/PhysRevB.106.235147} {\bibfield  {journal} {\bibinfo
   {journal} {Phys. Rev. B}\ }\textbf {\bibinfo {volume} {106}},\ \bibinfo
  {pages} {235147} (\bibinfo {year} {2022})}\BibitemShut {NoStop}%
\bibitem [{\citenamefont {Serbyn}\ \emph {et~al.}(2021)\citenamefont {Serbyn},
  \citenamefont {Abanin},\ and\ \citenamefont {Papi{\'c}}}]{Serbyn_2021}%
  \BibitemOpen
  \bibfield  {author} {\bibinfo {author} {\bibfnamefont {M.}~\bibnamefont
  {Serbyn}}, \bibinfo {author} {\bibfnamefont {D.~A.}\ \bibnamefont {Abanin}},\
  and\ \bibinfo {author} {\bibfnamefont {Z.}~\bibnamefont {Papi{\'c}}},\
  }\bibfield  {title} {\bibinfo {title} {Quantum many-body scars and weak
  breaking of ergodicity},\ }\href@noop {} {\bibfield  {journal} {\bibinfo
  {journal} {Nature Physics}\ }\textbf {\bibinfo {volume} {17}},\ \bibinfo
  {pages} {675} (\bibinfo {year} {2021})}\BibitemShut {NoStop}%
\bibitem [{\citenamefont {Moudgalya}\ \emph {et~al.}(2022)\citenamefont
  {Moudgalya}, \citenamefont {Bernevig},\ and\ \citenamefont
  {Regnault}}]{Moudgalya_2022_review}%
  \BibitemOpen
  \bibfield  {author} {\bibinfo {author} {\bibfnamefont {S.}~\bibnamefont
  {Moudgalya}}, \bibinfo {author} {\bibfnamefont {B.~A.}\ \bibnamefont
  {Bernevig}},\ and\ \bibinfo {author} {\bibfnamefont {N.}~\bibnamefont
  {Regnault}},\ }\bibfield  {title} {\bibinfo {title} {Quantum many-body scars
  and hilbert space fragmentation: a review of exact results},\ }\href
  {https://doi.org/10.1088/1361-6633/ac73a0} {\bibfield  {journal} {\bibinfo
  {journal} {Reports on Progress in Physics}\ }\textbf {\bibinfo {volume}
  {85}},\ \bibinfo {pages} {086501} (\bibinfo {year} {2022})}\BibitemShut
  {NoStop}%
\bibitem [{\citenamefont {Chandran}\ \emph {et~al.}(2023)\citenamefont
  {Chandran}, \citenamefont {Iadecola}, \citenamefont {Khemani},\ and\
  \citenamefont {Moessner}}]{Chandran_2023}%
  \BibitemOpen
  \bibfield  {author} {\bibinfo {author} {\bibfnamefont {A.}~\bibnamefont
  {Chandran}}, \bibinfo {author} {\bibfnamefont {T.}~\bibnamefont {Iadecola}},
  \bibinfo {author} {\bibfnamefont {V.}~\bibnamefont {Khemani}},\ and\ \bibinfo
  {author} {\bibfnamefont {R.}~\bibnamefont {Moessner}},\ }\bibfield  {title}
  {\bibinfo {title} {Quantum many-body scars: A quasiparticle perspective},\
  }\href@noop {} {\bibfield  {journal} {\bibinfo  {journal} {Annual Review of
  Condensed Matter Physics}\ }\textbf {\bibinfo {volume} {14}},\ \bibinfo
  {pages} {443} (\bibinfo {year} {2023})}\BibitemShut {NoStop}%
\bibitem [{\citenamefont {Bocini}\ and\ \citenamefont
  {Fagotti}(2024)}]{Bocini_2024}%
  \BibitemOpen
  \bibfield  {author} {\bibinfo {author} {\bibfnamefont {S.}~\bibnamefont
  {Bocini}}\ and\ \bibinfo {author} {\bibfnamefont {M.}~\bibnamefont
  {Fagotti}},\ }\bibfield  {title} {\bibinfo {title} {Growing schr\"odinger's
  cat states by local unitary time evolution of product states},\ }\href
  {https://doi.org/10.1103/PhysRevResearch.6.033108} {\bibfield  {journal}
  {\bibinfo  {journal} {Phys. Rev. Res.}\ }\textbf {\bibinfo {volume} {6}},\
  \bibinfo {pages} {033108} (\bibinfo {year} {2024})}\BibitemShut {NoStop}%
\bibitem [{\citenamefont {Weinberg}\ and\ \citenamefont
  {Bukov}(2017)}]{Quspin}%
  \BibitemOpen
  \bibfield  {author} {\bibinfo {author} {\bibfnamefont {P.}~\bibnamefont
  {Weinberg}}\ and\ \bibinfo {author} {\bibfnamefont {M.}~\bibnamefont
  {Bukov}},\ }\bibfield  {title} {\bibinfo {title} {Quspin: a python package
  for dynamics and exact diagonalisation of quantum many body systems part i:
  spin chains},\ }\href@noop {} {\bibfield  {journal} {\bibinfo  {journal}
  {SciPost Physics}\ }\textbf {\bibinfo {volume} {2}},\ \bibinfo {pages} {003}
  (\bibinfo {year} {2017})}\BibitemShut {NoStop}%
\bibitem [{\citenamefont {Gotta}\ \emph {et~al.}(2023)\citenamefont {Gotta},
  \citenamefont {Moudgalya},\ and\ \citenamefont {Mazza}}]{Gotta_2023}%
  \BibitemOpen
  \bibfield  {author} {\bibinfo {author} {\bibfnamefont {L.}~\bibnamefont
  {Gotta}}, \bibinfo {author} {\bibfnamefont {S.}~\bibnamefont {Moudgalya}},\
  and\ \bibinfo {author} {\bibfnamefont {L.}~\bibnamefont {Mazza}},\ }\bibfield
   {title} {\bibinfo {title} {Asymptotic quantum many-body scars},\ }\href
  {https://doi.org/10.1103/PhysRevLett.131.190401} {\bibfield  {journal}
  {\bibinfo  {journal} {Phys. Rev. Lett.}\ }\textbf {\bibinfo {volume} {131}},\
  \bibinfo {pages} {190401} (\bibinfo {year} {2023})}\BibitemShut {NoStop}%
\bibitem [{\citenamefont {Vardhan}\ and\ \citenamefont
  {Moudgalya}(2024)}]{Vardhan_2024}%
  \BibitemOpen
  \bibfield  {author} {\bibinfo {author} {\bibfnamefont {S.}~\bibnamefont
  {Vardhan}}\ and\ \bibinfo {author} {\bibfnamefont {S.}~\bibnamefont
  {Moudgalya}},\ }\bibfield  {title} {\bibinfo {title} {Entanglement dynamics
  from universal low-lying modes},\ }\href@noop {} {\bibfield  {journal}
  {\bibinfo  {journal} {arXiv preprint arXiv:2407.16763}\ } (\bibinfo {year}
  {2024})}\BibitemShut {NoStop}%
\end{thebibliography}%

\newpage
\clearpage
\renewcommand{\thefigure}{S\arabic{figure}}
\renewcommand{\thesection}{S\arabic{section}}
\renewcommand{\theequation}{S\arabic{equation}}
\renewcommand{\thepage}{S\arabic{page}}
\setcounter{page}{0}
\setcounter{section}{0}
\setcounter{secnumdepth}{2}

\onecolumngrid
\begin{center}
{\large \textbf{Online supplementary material for: \\ Open-system quantum many-body scars: a theory}}

\vspace{16pt}

Lorenzo Gotta$^{1}$

\begin{small}
$^1$\textit{ Department of Quantum Matter Physics, University of Geneva, 24 Quai Ernest-Ansermet, 1211 Geneva, Switzerland}
\end{small}

\vspace{10pt}

\today

\vspace{10pt}

\end{center}

%\section{Stationary operators and commutant}

%If one has $\mathcal{L}\rho =0$, then, by taking the scalar product with $\rho^{\dag}$, one gets:
%\begin{equation}
%    -iTr\left(\rho^{\dag}[\hat H,\rho]\right)-\frac{1}{2}Tr\left(\rho [\hat l_j,[\hat l_j,\rho]] \right) =0.
%\end{equation}
%The first term vanishes by cyclic invariance of the trace \textcolor{red}{(only on normal operators?)}, thus implying that the second term must equal zero. The latter vanishes iff $[\hat l_j,\rho]=0\,\,\forall j$ due to the frustration-free form of the dissipative part of the Lindbladian in case of hermitian jump operators. Thus, we have that $\mathcal{L}_d \rho = 0$. This implies that $0=\mathcal{L}\rho=\mathcal{L}_u \rho$, which means that $\mathcal{L}\rho=0$ implies $\mathcal{L}_u \rho=\mathcal{L}_d \rho = 0$. 
\section{Off-diagonal coherences between scar and thermal sectors}

Here, we present the calculation of the decay of off-diagonal coherences between the sector identified by a singlet $\ket{\psi}$ of the algebra $\mathcal{A}$ and its orthogonal complement, assuming that the latter is a Krylov subspace of $\mathcal{A}$ (we remark that the derivation remains valid as long as we consider the off-diagonal coherences between the one-dimensional sector spanned by a singlet of $\mathcal{A}$ and any other irrep of $\mathcal{A}$). Defining $\hat \Pi_{s} = \ket{\psi}\bra{\psi}$ and $\hat \Pi_{th} = 1 - \hat \Pi_s$, one needs to evaluate:
\begin{equation}\label{Eq:coh_off_diag}
    \frac{d}{dt} ||\hat\Pi_s e^{\mathcal{L}t}(\hat\rho_0) \hat\Pi_{th}||^2 = Tr\left[e^{\mathcal{L}t}\left(\mathcal{L}(\hat \Pi_{th}\hat\rho_0 \hat\Pi_s)\right) e^{\mathcal{L}t}\left(\hat \Pi_{s}\hat\rho_0 \hat\Pi_{th}\right) +e^{\mathcal{L}t}\left(\hat \Pi_{th}\hat\rho_0 \hat\Pi_{s}\right) e^{\mathcal{L}t}\left(\mathcal{L}(\hat \Pi_{s}\hat\rho_0 \hat\Pi_{th})\right) \right],
\end{equation}
where we used the fact that $\hat\Pi_s \mathcal{L}(\hat\rho_0)\hat\Pi_{th}=\mathcal{L}(\hat \Pi_s \hat\rho_0 \hat \Pi_{th})$ and, similarly, $\hat\Pi_{th} \mathcal{L}(\hat\rho_0)\hat\Pi_{s}=\mathcal{L}(\hat \Pi_{th} \hat\rho_0 \hat \Pi_{s})$, since $\hat\Pi_s$ and $\hat\Pi_{th}$, being projectors onto Krylov subspaces of $\mathcal{A}$, belong to its commutant algebra $\mathcal{C}$, and thus the off-diagonal coherences between the subspaces being the range of $\hat\Pi_s$ and $\hat\Pi_{th}$ are decoupled from the remainder of the operator space. We assume now the Lindblad superoperator to take the generic form $\mathcal{L}(\cdot)=-i\sum_{\alpha} g_{\alpha}[\hat h_{\alpha},\cdot]-\frac{1}{2}\sum_j [\hat l_j,[\hat l_j,\cdot]]$. Since the state $\ket{\psi}$ is a singlet of $\mathcal{A}$, and thus $\hat l_j \ket{\psi} = \lambda_j \ket{\psi}\,\,\forall j$ and $\hat h_{\alpha}\ket{\psi} = \epsilon_{\alpha}\ket{\psi}\,\,\forall \alpha$, one may observe that:
\begin{equation}\label{S:Eq:property_singlet}
    \begin{split}
    &\mathcal{L}(\hat\Pi_{th} \hat \rho_0 \hat\Pi_{s})= \hat H_{eff} \hat\Pi_{th} \hat \rho_0 \hat\Pi_{s}, \qquad  \mathcal{L}(\hat\Pi_{s} \hat \rho_0 \hat\Pi_{th})=  \hat\Pi_{s} \hat \rho_0 \hat\Pi_{th}\hat H_{eff}^{\dag},\\
    & \hat H_{eff} = -i \sum_{\alpha} J_{\alpha} (\hat h_{\alpha}-\epsilon_{\alpha})-\frac{1}{2}\sum_j \gamma_j (\hat l_j -\lambda_j)^2 = -i \hat H_1 -\frac{1}{2}\hat H_2,
    \end{split}
\end{equation}
Eq.~\eqref{Eq:coh_off_diag} can thus be rewritten as:
\begin{equation}
     \frac{d}{dt} ||\hat\Pi_s e^{\mathcal{L}t}(\hat\rho_0) \hat\Pi_{th}||^2 = Tr\left[e^{\mathcal{L}t}\left(\hat H_{eff} \hat\Pi_{th} \hat \rho_0 \hat\Pi_s \right) e^{\mathcal{L}t}\left( \hat\Pi_{s} \hat \rho_0 \hat\Pi_{th} \right) + h.c.  \right]. 
\end{equation}
Finally, exploiting once more the properties in Eq.\eqref{S:Eq:property_singlet} to perform the rewritings $e^{\mathcal{L}t}\left(\hat H_{eff} \hat\Pi_{th} \hat \rho_0 \hat\Pi_s \right)=e^{t\hat H_{eff}}\hat H_{eff} \hat\Pi_{th} \hat \rho_0 \hat\Pi_s $ and $e^{\mathcal{L}t}\left( \hat\Pi_{s} \hat \rho_0 \hat\Pi_{th} \right)= \hat\Pi_{s} \hat \rho_0 \hat\Pi_{th} e^{t\hat{H}_{eff}^{\dag}}$, together with the cyclic property of the trace, one obtains:
\begin{equation}\label{S:Eq:derivative_off_diag}
 \frac{d}{dt} ||\hat\Pi_s e^{\mathcal{L}t}(\hat\rho_0) \hat\Pi_{th}||^2 = -Tr\left[ \hat\Pi_{s} \hat \rho_0 \hat\Pi_{th}\, e^{t\hat H_{eff}^{\dag}}\sum_j \gamma_j (\hat l_j -\lambda_j)^2\, e^{t\hat H_{eff}}\, \hat\Pi_{th} \hat \rho_0 \hat\Pi_{s}  \right].
\end{equation}

As an application of Eq.\eqref{S:Eq:derivative_off_diag}, one may observe that, introducing the shorthand notation $e^{\mathcal{L}t}(\hat\rho_0)= \hat\rho(t)$ and assuming that the Hamiltonian $\hat H_2$ satisfies $\hat \Pi_{th} \hat H_2 \hat \Pi_{th}\geq g > 0$, which is equivalent to asking that all of its eigenstates in $\Pi_{th}$ possess a finite energy bounded from below by a positive constant for any value of the size $L$, i.e., that $\hat H_2$ is a gapped Hamiltonian, one obtains finally the bound:
\begin{equation}
 \frac{d}{dt} ||\hat\Pi_s e^{\mathcal{L}t}(\hat\rho_0) \hat\Pi_{th}||^2 \leq -g ||\hat\Pi_s e^{\mathcal{L}t}(\hat\rho_0) \hat\Pi_{th}||^2 \implies ||\hat\Pi_s e^{\mathcal{L}t}(\hat\rho_0) \hat\Pi_{th}||^2 \leq ||\hat\Pi_s \hat\rho_0 \hat\Pi_{th}||^2 e^{-gt}.
\end{equation}

\section{Irreducibility of the spin-$1/2$ algebras with isolated OSQMBS}

\subsection{Model 1}

Let us prove the irreducibility of the algebra $\mathcal{A}_{1,\dots,L}=\langle\langle \{\hat\sigma_j^x \hat \sigma_{j+1}^x + \hat \sigma_j^y \hat \sigma_{j+1}^y \}_{j=1}^{L-1}, \{\hat\sigma_j^x (1-\hat\sigma_{j+1}^z)/2 \}_{j=1}^{L-1}, (1-\hat\sigma_{j}^z)\hat\sigma_{j+1}^x/2 \}_{j=1}^{L-1} \rangle\rangle$ in the orthogonal complement to $W_{scar,L} := span \{\ket{F,\uparrow}\}$. In order to achieve such a goal, we need to show that, given an orthonormal basis $\{\ket{\psi_{\alpha}}:\alpha=1,\dots,2^L-1 \}$ of $W_{scar,L}^{\perp}$, then $\ket{\psi_{\alpha}}\bra{\psi_{\beta}}\in \mathcal{A}_{1,\dots,L}\,\,\forall \alpha,\beta$.

\paragraph{Induction basis $L=2$:} 

we start the proof by considering the case of $L=2$ sites, labelled generically $j,\, j+1$. By noticing that $\hat\sigma_j^z(\hat\sigma_j^+ \hat\sigma_{j+1}^- +h.c.)=\hat\sigma_j^+ \hat\sigma_{j+1}^- -\hat\sigma_j^- \hat\sigma_{j+1}^+$ belongs to $\mathcal{A}$ and adding it to the exchange term $\hat\sigma_j^+ \hat\sigma_{j+1}^- +h.c.$, we immediately obtain that both $\hat\sigma_j^+\hat\sigma_{j+1}^-$ and its hermitian conjugate belong to $\mathcal{A}$. Moreover, introducing the notation $\hat P_{j,\downarrow}=(1-\hat\sigma_j^z)/2$ and $\hat P_{j,\uparrow}=(1+\hat\sigma_j^z)/2$, we observe that $(1+\hat\sigma_j^z) \sigma_j^x (1-\hat\sigma_j^z)/4 = \hat\sigma_j^+ \hat P_{j+1,\downarrow}$ belongs to $\mathcal{A}$, as well as its hermitian conjugate. The same holds for $(1-\hat\sigma_j^z)\hat\sigma_{j+1}^x (1+\hat\sigma_{j+1}^z)/4 = \hat P_{j,\downarrow}\hat\sigma_{j+1}^-$ and its hermitian conjugate. Finally, $(\hat\sigma_j^+ \hat\sigma_{j+1}^-)(\hat\sigma_j^- \hat\sigma_{j+1}^+)=\hat P_{j,\uparrow}\hat P_{j+1,\downarrow}$, $(\hat\sigma_j^- \hat\sigma_{j+1}^+)(\hat\sigma_{j}^+ \hat\sigma_{j+1}^-)=\hat P_{j,\downarrow} \hat P_{j+1,\uparrow}$ and $\left[\hat\sigma_j^x \hat P_{j+1,\downarrow}\right]^2 \left[\hat P_{j,\downarrow}\hat\sigma_{j+1}^x \right]^2=\hat P_{j,\downarrow} \hat P_{j+1,\downarrow}$ belong to $\mathcal{A}$. This completes the proof for $L=2$.

\paragraph{Induction step:}

Assuming that the algebra $\mathcal{A}_{1,\dots,k}$ is irreducible for any choice of $2\leq k \leq L$ sites over which it is defined, we are going to show that is is irreducible also when defined over $L+1$ sites,i.e., for $k=L+1$. By the induction hypothesis, the algebras $\mathcal{A}_{1,\dots,L}$ and $\mathcal{A}_{2,\dots,L+1}$ contain $\ket{\psi_{\alpha}} \otimes_{j=1}^{L}\bra{\downarrow}_j$ for every element of an orthonormal basis $\{\ket{\psi_{\alpha}}:\alpha=1,\dots,2^L-1\}$ of $W_{scar,L}^{\perp}$ over the $L$ sites that they are defined over. The above statement is true in particular when $L=2$. Thus, the following compositions of operators belong to $\mathcal{A}_{1,\dots,L+1}$:
\begin{equation}\label{Eq:induction0}
    \begin{split}
      & \left(\ket{\psi_{\alpha}}_{1,\dots,L}\otimes_{j=1}^{L}\bra{\downarrow}_j\right)\left(\ket{\downarrow}_{L} \ket{\downarrow}_{L+1} \bra{\downarrow}_L\bra{\downarrow}_{L+1}\right) =\ket{\psi_{\alpha}}_{1,\dots,L}\ket{\downarrow}_{L+1}\otimes_{j=1}^{L+1}\bra{\downarrow}_j\\
      & \left(\ket{\psi_{\alpha}}_{1,\dots,L}\otimes_{j=1}^{L}\bra{\downarrow}_j\right) \left(\ket{\downarrow}_{L} \ket{\uparrow}_{L+1} \bra{\downarrow}_{L}\bra{\downarrow}_{L+1}\right) =\ket{\psi_{\alpha}}_{1,\dots,L}\ket{\uparrow}_{L+1}\otimes_{j=1}^{L+1}\bra{\downarrow}_j\\
      &\left(\ket{\psi_{\alpha}}_{2,\dots,L+1}\otimes_{j=2}^{L+1}\bra{\downarrow}_j\right) \left(\ket{\uparrow}_{1} \ket{\downarrow}_2 \bra{\downarrow}_{1}\bra{\downarrow}_2 \right) =\ket{\uparrow}_{1}\ket{\psi_{\alpha}}_{2,\dots,L+1}\otimes_{j=1}^{L+1}\bra{\downarrow}_j\\
      &\left(\ket{\psi_{\alpha}}_{2,\dots,L+1}\otimes_{j=2}^{L+1}\bra{\downarrow}_j\right) \left(\ket{\downarrow}_{1} \ket{\downarrow}_2 \bra{\downarrow}_{1}\bra{\downarrow}_2\right) =\ket{\downarrow}_{1}\ket{\psi_{\alpha}}_{2,\dots,L+1}\otimes_{j=1}^{L+1}\bra{\downarrow}_j.
    \end{split}
\end{equation}
    Since we have shown that r.h.s. of Eq.\eqref{Eq:induction0} belongs to $\mathcal{A}_{1,\dots,L+1}$ and this fact implies that $\ket{\varphi_{\alpha}}\otimes_{j=1}^{L+1}\bra{\downarrow}_j\in \mathcal{A}_{1,\dots,L+1}$ for every element $\ket{\varphi_{\alpha}}$ of an orthonormal basis of $W_{scar,L+1}^{\perp}$, we can conclude that $\mathcal{A}_{1,\dots,L+1}$ is indeed irreducible within $W_{scar,L+1}^{\perp}$, thus completing the proof.

\subsection{Model 2}

Let us consider the algebra $\mathcal{A}_{1,\dots,L}=\langle\langle\{\hat \sigma_j^x \hat \sigma_{j+1}^x+\hat\sigma_j^y \hat\sigma_{j+1}^y\}_{j=1}^{L-1}, \{\hat \sigma_j^z \}_{j=1}^L, \{\hat\sigma_j^+ \hat \sigma_{j+1}^+ \hat \sigma_{j+2}^- +h.c. \}_{j=1}^{L-2} \rangle\rangle$ and let us prove its irreducibility in the orthogonal complement to $W_{scar,L}:= span\{\ket{F\uparrow},\,\ket{F\downarrow} \}$.
To this end, we need to show that, given an orthonormal basis $\{\ket{\psi_{\alpha}}:\alpha=1,\dots,2^L-2 \}$ of $W_{scar,L}^{\perp}$, then $\ket{\psi_{\alpha}}\bra{\psi_{\beta}}\in \mathcal{A}_{1,\dots,L}\,\,\forall \alpha,\beta$. 

\paragraph{Induction basis $L=3$:}
considering the case of a system with $L=3$ sites labelled $j,\,j+1,\,j+2$, we start by taking the product $\hat\sigma_j^z (\hat \sigma_j^+\hat\sigma_{j+1}^- + h.c.)=\hat \sigma_j^+\hat\sigma_{j+1}^- - h.c.$, and both summing and subtracting the result to the exchange term $\hat \sigma_j^+\hat\sigma_{j+1}^- + h.c.$. Thus, we get that both $\hat\sigma_j^+ \hat\sigma_{j+1}^-$ and its hermitian conjugate belong to the algebra. By introducing the notation $\hat P_{j,\uparrow} = \ket{\uparrow}_j \bra{\uparrow}_j,\, \hat P_{j,\downarrow} = \ket{\downarrow}_j \bra{\downarrow}_j$ and taking their products in both orders, one gets that $\hat P_{j,\uparrow}\hat P_{j+1,\downarrow}$ and $\hat P_{j,\downarrow}\hat P_{j+1,\uparrow}$ belong to the algebra as well. Similar results hold by translating $j\rightarrow j+1$ in the above formulas. Therefore, all the products $\hat P_{j,\uparrow} \hat P_{j+1,\downarrow} \hat P_{j+2,\uparrow}$, $\hat P_{j,\downarrow}\hat P_{j+1,\uparrow} \hat P_{j+2,\downarrow}$, $(\hat\sigma_j^+ \hat\sigma_{j+1}^-) (\hat\sigma_{j+1}^+ \hat\sigma_{j+2}^-)= \hat\sigma_{j}^+ \hat P_{j+1,\downarrow} \hat\sigma_{j+2}^-$, $(\hat\sigma_j^- \hat\sigma_{j+1}^+)(\hat\sigma_{j+1}^- \hat\sigma_{j+2}^+)=\hat\sigma_j^- \hat P_{j+1,\uparrow} \hat\sigma_{j+2}^+$, $(\hat\sigma_j^+ \hat\sigma_{j+1}^-)(\hat P_{j+1,\uparrow} \hat P_{j+2,\downarrow})=\hat\sigma_j^+ \hat\sigma_{j+1}^- \hat P_{j+2,\downarrow}$, $(\hat\sigma_j^- \hat\sigma_{j+1}^+)(\hat P_{j+1,\downarrow}\hat P_{j+2,\uparrow})= \hat\sigma_j^- \hat\sigma_{j+1}^+ \hat P_{j+2,\uparrow}$, $(\hat P_{j,\downarrow} \hat P_{j+1,\uparrow})(\hat\sigma_{j+1}^+ \hat\sigma_{j+2}^-)=\hat P_{j,\downarrow} \hat\sigma_{j+1}^+ \hat\sigma_{j+2}^-$, $(\hat P_{j,\uparrow} \hat P_{j+1,\downarrow})(\hat\sigma_{j+1}^- \hat\sigma_{j+2}^+)=\hat P_{j,\uparrow} \hat\sigma_{j+1}^- \hat\sigma_{j+2}^+$, and their hermitian conjugates, belong to the algebra.
Moreover, exploiting $\hat\sigma_j^z (\hat\sigma_j^+ \hat\sigma_{j+1}^+ \hat\sigma_{j+2}^- +h.c.)= \hat\sigma_j^+ \hat\sigma_{j+1}^+ \hat\sigma_{j+2}^- -h.c.$ and both adding and subtracting this term to the anomalous exchange term $\hat\sigma_j^+ \hat\sigma_{j+1}^+ \hat\sigma_{j+2}^- +h.c.$, one obtains that both $\hat\sigma_j^+ \hat\sigma_{j+1}^+ \hat\sigma_{j+2}^-$ and its hermitian conjugate $\hat\sigma_j^- \hat\sigma_{j+1}^- \hat\sigma_{j+2}^+ $ belong to the algebra. By taking the product between the two in both orders, one obtains further that $\hat P_{j,\uparrow} \hat P_{j+1,\uparrow}\hat P_{j+2,\downarrow}$ and $\hat P_{j,\downarrow} \hat P_{j+1,\downarrow} \hat P_{j+2,\uparrow}$ belong to the algebra. Moreover, we know that $\hat \sigma_j^+ \hat P_{j+1,\downarrow} \hat\sigma_{j+2}^- +\hat \sigma_j^+ \hat P_{j+1,\uparrow} \hat\sigma_{j+2}^- = \hat\sigma_j^+ \hat\sigma_{j+2}^-$ and its hermitian conjugate belong to the algebra. Hence, we can write that $(\hat\sigma_{j+1}^- \hat\sigma_{j+2}^+)(\hat\sigma_j^+\hat\sigma_{j+1}^+\hat\sigma_{j+2}^-)=\hat\sigma_j^+\hat P_{j+1,\downarrow}\hat P_{j+2,\uparrow}$, $(\hat\sigma_{j}^- \hat\sigma_{j+2}^+)(\hat\sigma_j^+\hat\sigma_{j+1}^+\hat\sigma_{j+2}^-)=\hat P_{j,\downarrow}\hat \sigma_{j+1}^+\hat P_{j+2,\uparrow}$, $(\hat\sigma_{j}^+ \hat\sigma_{j+2}^-)(\hat\sigma_j^-\hat\sigma_{j+1}^-\hat\sigma_{j+2}^+)=\hat P_{j,\uparrow}\hat \sigma_{j+1}^-\hat P_{j+2,\downarrow}$, $(\hat\sigma_{j+1}^+ \hat\sigma_{j+2}^-)(\hat\sigma_j^-\hat\sigma_{j+1}^-\hat\sigma_{j+2}^+)=\hat \sigma_{j}^- \hat P_{j+1,\uparrow}\hat P_{j+2,\downarrow}$, $(\hat P_{j,\downarrow}\hat\sigma_{j+1}^+ \hat P_{j+2,\uparrow})(\hat\sigma_{j+1}^- \hat\sigma_{j+2}^+)=\hat P_{j,\downarrow}\hat P_{j+1,\uparrow}\hat\sigma_{j+2}^+$, $(\hat \sigma_{j}^+\hat P_{j+1,\downarrow}^+ \hat P_{j+2,\uparrow})(\hat\sigma_{j}^- \hat\sigma_{j+2}^+)=\hat P_{j,\uparrow}\hat P_{j+1,\downarrow}\hat\sigma_{j+2}^+$ and their hermitian conjugates belong to the algebra. One can see as well that $(\hat P_{j,\downarrow}\hat \sigma_{j+1}^+\hat P_{j+2,\uparrow})(\hat P_{j,\downarrow}\hat \sigma_{j+1}^-\hat P_{j+2,\uparrow})=\hat P_{j,\downarrow}\hat P_{j+1,\uparrow}\hat P_{j+2,\uparrow}$ and $(\hat P_{j,\uparrow}\hat \sigma_{j+1}^-\hat P_{j+2,\downarrow})(\hat P_{j,\uparrow}\hat \sigma_{j+1}^+\hat P_{j+2,\downarrow})=\hat P_{j,\uparrow}\hat P_{j+1,\downarrow}\hat P_{j+2,\downarrow}$ belong to the algebra. Finally, one observes that $(\hat\sigma_j^+ \hat\sigma_{j+1}^-)(\hat P_{j,\downarrow} \hat P_{j+1,\uparrow} \hat\sigma_{j+2}^-)=\hat\sigma_j^+ \hat\sigma_{j+1}^- \hat\sigma_{j+2}^-$, $(\hat\sigma_j^+ \hat\sigma_{j+1}^-)(\hat P_{j,\downarrow} \hat P_{j+1,\uparrow} \hat\sigma_{j+2}^+)=\hat\sigma_j^+ \hat\sigma_{j+1}^- \hat\sigma_{j+2}^+$, $(\hat\sigma_j^- \hat\sigma_{j+1}^+)(\hat P_{j,\uparrow} \hat P_{j+1,\downarrow} \hat\sigma_{j+2}^+)=\hat\sigma_j^- \hat\sigma_{j+1}^+ \hat\sigma_{j+2}^+$ and their hermitian conjugates belong to the algebra as well. In this way, we have realized the $6\times 6=36$ ket-bra operators acting in the space of one- and two-spin-flip states ($3+3$ states in total). 

\paragraph{Induction step:}
Assuming that the algebra $\mathcal{A}_{1,\dots,k}$ is irreducible for any choice of $3\leq k \leq L$ sites over which it is defined, we are going to show that is is irreducible also when defined over $L+1$ sites. By the induction hypothesis, the algebras $\mathcal{A}_{1,\dots,L}$ and $\mathcal{A}_{2,\dots,L+1}$ contain $\ket{\psi_{\alpha}}\bra{\downarrow}_1 \otimes_{j=2}^{L-1}\bra{\uparrow}_j\bra{\downarrow} $ for every element of an orthonormal basis $\{\ket{\psi_{\alpha}}:\alpha=1,\dots,2^L-2\}$ of $W_{scar}^{\perp}$ over the $L$ sites that they are defined over. The above statement is true in particular when $L=3$. Thus, the following compositions of operators belong to $\mathcal{A}_{1,\dots,L+1}$:
\begin{equation}\label{Eq:induction}
    \begin{split}
      & \left(\ket{\psi_{\alpha}}_{1,\dots,L}\bra{\downarrow}_1 \otimes_{j=2}^{L-1}\bra{\uparrow}_j \bra{\downarrow}_L\right)\left(\ket{\uparrow}_{L-1} \ket{\downarrow}_L \ket{\downarrow}_{L+1}\bra{\uparrow}_{L-1}\bra{\uparrow}_L\bra{\downarrow}_{L+1}\right) =\ket{\psi_{\alpha}}_{1,\dots,L}\ket{\downarrow}_{L+1}\bra{\downarrow}_1 \otimes_{j=2}^{L}\bra{\uparrow}_j \bra{\downarrow}_{L+1}\\
      & \left(\ket{\psi_{\alpha}}_{1,\dots,L}\bra{\downarrow}_1 \otimes_{j=2}^{L-1}\bra{\uparrow}_j \bra{\downarrow}_L\right) \left(\ket{\uparrow}_{L-1} \ket{\downarrow}_L \ket{\uparrow}_{L+1}\bra{\uparrow}_{L-1}\bra{\uparrow}_L\bra{\downarrow}_{L+1}\right) =\ket{\psi_{\alpha}}_{1,\dots,L}\ket{\uparrow}_{L+1}\bra{\downarrow}_1 \otimes_{j=2}^{L}\bra{\uparrow}_j \bra{\downarrow}_{L+1}\\
      &\left(\ket{\psi_{\alpha}}_{2,\dots,L+1}\bra{\downarrow}_2 \otimes_{j=3}^{L}\bra{\uparrow}_j \bra{\downarrow}_{L+1}\right) \left(\ket{\downarrow}_{1} \ket{\downarrow}_2 \ket{\uparrow}_{3}\bra{\downarrow}_{1}\bra{\uparrow}_2\bra{\uparrow}_{3}\right) =\ket{\downarrow}_{1}\ket{\psi_{\alpha}}_{2,\dots,L+1}\bra{\downarrow}_1 \otimes_{j=2}^{L}\bra{\uparrow}_j \bra{\downarrow}_{L+1}\\
      &\left(\ket{\psi_{\alpha}}_{2,\dots,L+1}\bra{\downarrow}_2 \otimes_{j=3}^{L}\bra{\uparrow}_j \bra{\downarrow}_{L+1}\right) \left(\ket{\uparrow}_{1} \ket{\downarrow}_2 \ket{\uparrow}_{3}\bra{\downarrow}_{1}\bra{\uparrow}_2\bra{\uparrow}_{3}\right) =\ket{\uparrow}_{1}\ket{\psi_{\alpha}}_{2,\dots,L+1}\bra{\downarrow}_1 \otimes_{j=2}^{L}\bra{\uparrow}_j \bra{\downarrow}_{L+1}.
    \end{split}
\end{equation}
    Since we have shown that r.h.s. of Eq.\eqref{Eq:induction} belongs to $\mathcal{A}_{1,\dots,L+1}$ and this fact implies that $\ket{\varphi_{\alpha}}\bra{\downarrow}_1 \otimes_{j=2}^{L}\bra{\uparrow}_j \bra{\downarrow}_{L+1}\in \mathcal{A}_{1,\dots,L+1}$ for every element $\ket{\varphi_{\alpha}}$ of an orthonormal basis of $W_{scar,L+1}^{\perp}$, we can conclude that $\mathcal{A}_{1,\dots,L+1}$ is indeed irreducible within $W_{scar,L+1}^{\perp}$, thus completing the proof.

\section{OSAQMBS and their dynamical signatures}

\subsection{Fidelity dynamics}

In this subsection, we consider the short-time expansion of the fidelity of the state  $\hat\rho_{n,k}=\ket{n,k}\bra{n,k}$, defined as:
\begin{equation}\label{Eq:fidelity}
    \mathcal{F}(t)= Tr[\hat\rho_{n,k}e^{t\mathcal{L}}(\hat\rho_{n,k})],
\end{equation}
where the Lindbladian generator is realized by means of the bond algebra:
\begin{equation}\label{S:Eq:bond_algebra_tower}
     \mathcal{A} = \langle\langle\{\hat S_j^x \hat S_{j+1}^x+\hat S_j^y \hat S_{j+1}^y\},\{ (\hat S_j^z)^2 \},
        \{(\hat S_j^z + \hat S_{j+1}^z)(1-\hat S_j^z \hat S_{j+1}^z)\}, \hat S_{tot}^z  \rangle\rangle ,
\end{equation}
Since the adjoint action of all generators but the set $\{\hat S_j^x \hat S_{j+1}^x+\hat S_j^y \hat S_{j+1}^y\}$, we highlight two different possible instances: the Lindbladians in which the latter terms are present only in the generator of the unitary evolution, and the mirror case in which they appear solely in the dissipative generator of the evolution.

We start by treating the latter case, in which case the short-time first-order correction to the fidelity is non-vanishing and reads:
\begin{equation}
    \mathcal{F}(t) = 1 + \frac{d}{dt}\mathcal{F}(t)\Bigl|_{t=0}\, t + O(t^2),
\end{equation}
where the first time-derivative of Eq.~\eqref{Eq:fidelity} at time $t=0$ results in the expression:
\begin{equation}\label{Eq:fid_calc}
    \begin{split}
      \frac{d}{dt}\mathcal{F}(t)|_{t=0} = Tr[\hat\rho_{n,k}\mathcal{L}(\hat\rho_{n,k})]=-\frac{1}{2}\sum_j \gamma_j Tr\Bigl[\hat\rho_{n,k}[\hat l_j,[\hat l_j,\hat\rho_{n,k}]]\Bigr]= -\sum_j \gamma_j \left[\bra{n,k}\hat l_j^2 \ket{n,k}-\left(\bra{n,k}\hat l_j \ket{n,k} \right)^2 \right],  
    \end{split}
\end{equation}
where we denoted $\hat l_j = \hat S_j^x \hat S_{j+1}^x+\hat S_j^y \hat S_{j+1}^y$.
In order to make progress, we make use of the results:
\begin{equation} \label{Eq:aqmbs_res}
    \begin{split}
        & \bra{n,k} \hat l_j \ket{n,k} =0,\\
        & \bra{n,k} \hat l_j^2 \ket{n,k} = \frac{4\cos^2 (k/2)}{\mathcal{N}_{n,k}} \frac{\binom{L-2}{n-1}}{\binom{L}{n-1}} = \frac{4}{L}\cos^2\left(\frac{k}{2}\right) \isEquivTo{L\gg 1} \frac{2}{L} (k-\pi)^2 = \frac{8\pi^2}{L^3},
    \end{split}
\end{equation}
where we used the fact that $k=\pi \pm \frac{2\pi}{L}$, and of course assumed $n<L$, since the expectation value of $\hat l_j^2$ is otherwise trivially equal to zero.

The expression in Eq.~\eqref{Eq:fid_calc} can thus be simplified to:
\begin{equation}
   \frac{d}{dt}\mathcal{F}(t)|_{t=0} =  -\frac{4}{L}\cos^2\left(\frac{k}{2}\right)\sum_j \gamma_j,  
\end{equation}
which, in the large-$L$ limit, leads to the rewriting:
\begin{equation}\label{S:Eq:scaling_diss}
   \frac{d}{dt}\mathcal{F}(t)|_{t=0} = -  \frac{8\pi^2}{L^2}  \tilde{\gamma},
\end{equation}
where $\tilde{\gamma}=\sum_j \gamma_j /L$ is the sample average of the decay rates $\{\gamma_j\}$ and which indeed vanishes in the thermodynamic limit. Moreover, the scaling highlighted in Eq.\eqref{S:Eq:scaling_diss} shows that the first-order correction to the fidelity scales as $t/L^2$, justifying the Ansatz employed to analyze the numerical data in the main text.

In the opposite case, when the exchange term appears only in the dissipative terms of the Lindbladian, we need to consider the second-order correction, as the first-order one vanishes. The short-time expansion of the fidelity now reads:
\begin{equation}
    \mathcal{F}(t) = 1 + \frac{1}{2}\frac{d^2}{dt^2}\mathcal{F}(t)\Bigl|_{t=0}\, t^2 + O(t^3),
\end{equation}
where, introducing the notation $\hat h_j =\hat S_j^x \hat S_{j+1}^x+\hat S_j^y \hat S_{j+1}^y $ and their coefficients in the Hamiltonian within the generator of the unitary evolution as $\{ g_j\}$:
\begin{equation}\label{S:Eq:second_derivative}
    \frac{d^2}{dt^2}\mathcal{F}(t)\Bigl|_{t=0} = Tr[\hat\rho_{n,k} \mathcal{L}^2 (\hat\rho_{n,k})] = -2\sum_{j} g_j^2 \left( \bra{\psi_{n,k}} \hat h_j^2 \ket{\psi_{n,k}} -\bra{\psi_{n,k}} \hat h_j \ket{\psi_{n,k}}^2  \right).
\end{equation}
In deriving the above result, we employed the first line of Eq.\eqref{Eq:aqmbs_res}, as well as the fact that $\bra{\psi_{n,k}} \hat h_j \hat h_l \ket{\psi_{n,k}}=0$ when $j\neq l$. Using the second line of Eq.\eqref{Eq:aqmbs_res}, we rewrite Eq.\eqref{S:Eq:second_derivative} as:
\begin{equation}
 Tr[\hat\rho_{n,k} \mathcal{L}^2 (\hat\rho_{n,k})] = -2\cdot  \frac{4}{L}\cos^2\left(\frac{k}{2}\right)\sum_{j} g_j^2  = -\frac{16 \pi^2}{L^2} \tilde{g^2},
\end{equation}
where we introduced the notation $\tilde{g^2}= \frac{1}{L}\sum_j g_j^2$. The above expression shows that the short-time expansion of the fidelity depends on the ratio $t^2 /L^2$, motivating the Ansatz employed in the main text to interpret the numerical data, on top of showing a signature of dynamical slowdown as $L\rightarrow +\infty$.

\subsection{Observables' dynamics}

We continue our treatment by discussing the scaling of the first-order correction in time $t$ to the initial expected value of a generic local observable $\hat O_{l}$, supported in a local neighborhood around site $l$ and with $||\hat O_l||\sim O(1)$, over the state $\hat\rho_{n,k}=\ket{n,k}\bra{n,k}$, which we manipulate as follows:
\begin{equation}
    \begin{split}
        & \frac{d}{dt}\langle \hat O_l \rangle(t)|_{t=0} = Tr[\hat O_l \mathcal{L}(\hat\rho_{n,k})]=-i\sum_{\alpha} g_{\alpha}\left(\hat O_l, [\hat h_{\alpha},\hat\rho_{n,k}]\right)-\frac{1}{2}\sum_j \gamma_j \left(\hat O_l, [\hat l_j,[\hat l_j,\hat\rho_{n,k}]] \right),
    \end{split}
\end{equation}
where we introduced the operator-space scalar product $(\hat A, \hat B):= Tr[\hat A^{\dag} \hat B]$ and the Lindbladian is realized through the bond algebra introduced in Eq.\eqref{S:Eq:bond_algebra_tower}. We first notice that the only set of local terms, among the generators of $\mathcal{A}$ in Eq.\eqref{S:Eq:bond_algebra_tower}, whose adjoint action on the state $\hat\rho_{n,k}$ is nonvanishing is given by $\{\hat S_j^x \hat S_{j+1}^x+\hat S_j^y \hat S_{j+1}^y\}$; therefore, by introducing the notation $\hat h_j=\hat l_j := \hat S_j^x \hat S_{j+1}^x+\hat S_j^y \hat S_{j+1}^y$, we are able to get the following bound on $Tr[\hat O_l \mathcal{L}(\hat\rho_{n,k})]$:
\begin{equation}\label{Eq:obs_calc}
    \begin{split}
    &\Bigl|-i\sum_{j} g_{j}\left(\hat O_l, [\hat h_{j},\hat\rho_{n,k}]\right)-\frac{1}{2}\sum_j \gamma_j \left(\hat O_l, [\hat l_j,[\hat l_j,\hat\rho_{n,k}]] \right)\Bigr| \leq ||\hat O_l||\left(\sum_{j} |g_{j}| \, \Bigl|\Bigl|[\hat h_{j},\hat\rho_{n,k}]\,\Bigr|\Bigr| +\frac{1}{2}\sum_j \gamma_j \Bigl|\Bigl|[\hat l_j,[\hat l_j,\hat\rho_{n,k}]]\Bigr|\Bigr| \right)=\\
    &=||\hat O_l||\left(\sum_{j} |g_{j}|\sqrt{2\bra{n,k}\hat h_{j}^2\ket{n,k}-2\bra{n,k}\hat h_{j}\ket{n,k}^2} +\frac{1}{2}\sum_j \gamma_j \sqrt{2\bra{n,k}\hat l_j^4 \ket{n,k}+6 \bra{n,k}\hat l_j^2 \ket{n,k}^2}\right)\leq \\
    & \leq ||\hat O_l||\left(\sum_{j} |g_{j}|\sqrt{2\bra{n,k}\hat h_{j}^2\ket{n,k}-2\bra{n,k}\hat h_{j}\ket{n,k}^2} +\frac{1}{2}\sum_j \gamma_j \sqrt{2||\hat l_j||^2\bra{n,k}\hat l_j^2 \ket{n,k}+6 \bra{n,k}\hat l_j^2 \ket{n,k}^2}\right),
    \end{split}
\end{equation}
where we used the fact that $\bra{n,k}\hat l_j^4 \ket{n,k}=||\hat l_j^2 \ket{n,k}||^2 \leq ||\hat l_j||^2 ||\hat l_j \ket{n,k}||^2= ||\hat l_j||^2 \bra{n,k}\hat l_j^2 \ket{n,k}$ and that $\bra{n,k} \hat l_j \ket{n,k}=0$.
Plugging in the final expression in Eq.~\eqref{Eq:obs_calc} the results shown in Eq.~\eqref{Eq:aqmbs_res} and observing that $C:=||\hat l_j||$ does not depend on $j$, one arrives to the final bound:
\begin{equation}
    \begin{split}
      \Bigl|\frac{d}{dt}\langle \hat O_l \rangle(t)\Bigl|_{t=0}\Bigr| \leq ||\hat O_l||\left(\frac{4\pi}{\sqrt{L}}\max_{j} |g_{j}| +\frac{2\pi C}{\sqrt{L}} \sqrt{1+\frac{24\pi^2}{C^2 L^3}} \max_j\gamma_j \right),
    \end{split}
\end{equation}
where the r.h.s. has already been expressed in the limit of large size $L$, which clearly shows that the upper bound tends to zero as $L\rightarrow +\infty$.

\subsection{Decay of off-diagonal coherences}

Now, we consider the quantity defined in Eq.\eqref{Eq:coh_off_tower}; keeping in mind that the bond algebra $\mathcal{A}$ of the system reads:
\begin{equation}
     \mathcal{A} = \langle\langle\{\hat S_j^x \hat S_{j+1}^x+\hat S_j^y \hat S_{j+1}^y\},\{ (\hat S_j^z)^2 \},
        \{(\hat S_j^z + \hat S_{j+1}^z)(1-\hat S_j^z \hat S_{j+1}^z)\}, \hat S_{tot}^z  \rangle\rangle ,
\end{equation}
we assume that the dissipative jump operators are given by the exchange terms, i.e., $\hat l_j = \hat S_j^x \hat S_{j+1}^x+\hat S_j^y \hat S_{j+1}^y$, while all the families of operators appearing as generators of $\mathcal{A}$ appear in the generator of the unitary evolution. Then, using Eq.\eqref{S:Eq:property_singlet}, it is easy to perform the following manipulation:
\begin{equation}\label{Eq:exp_off_coh}
\begin{split}
    &||\hat\Pi_{W_{n_0}}e^{t\mathcal{L}}(\hat\rho_0)\ket{\psi_{n_0}}\bra{\psi_{n_0}}||^2= Tr\Bigl[e^{t\mathcal{L}}(\ket{\psi_{n_0}}\bra{\psi_{n_0}}\hat\rho_0\hat\Pi_{W_{n_0}})e^{t\mathcal{L}}(\hat\Pi_{W_{n_0}}\hat\rho_0\ket{\psi_{n_0}}\bra{\psi_{n_0}})\Bigr]=\\
    & Tr\Bigl[\ket{\psi_{n_0}}\bra{\psi_{n_0}}\hat\rho_0\hat\Pi_{W_{n_0}}e^{it\hat H_1-\frac{t}{2}\hat H_2} e^{-it\hat H_1-\frac{t}{2}\hat H_2}\hat\Pi_{W_{n_0}}\hat\rho_0\ket{\psi_{n_0}}\bra{\psi_{n_0}}\Bigr]=\frac{1}{4}\bra{n_0,k}e^{it\hat H_1-\frac{t}{2}\hat H_2} e^{-it\hat H_1-\frac{t}{2}\hat H_2}\ket{n_0,k},
\end{split}
\end{equation}
where we used the notation introduced in Eq.\eqref{S:Eq:property_singlet} and, in the last equality, we have exploited the fact that the initial state $\hat \rho_0$ has the form introduced in Eq.~\eqref{Eq:initial_state}, so that it is easy to see that the relation $\ket{\psi_{n_0}}\bra{\psi_{n_0}}\hat\rho_0\hat\Pi_{W_{n_0}}=\frac{1}{2}\ket{\psi_{n_0}}\bra{n_0,k}$ is satisfied. Expanding Eq.\eqref{Eq:exp_off_coh} to first-order in time $t$, one finally obtains:
\begin{equation}\label{Eq:off_diag_coh_calc}
||\hat\Pi_{W_{n_0}}e^{t\mathcal{L}}(\hat\rho_0)\ket{\psi_{n_0}}\bra{\psi_{n_0}}||^2= \frac{1}{4}\left[1-t\sum_j\gamma_j \bra{n_0,k} \left(\hat S_j^x \hat S_{j+1}^x+\hat S_j^y \hat S_{j+1}^y\right)^2 \ket{n_0,k} +O(t^2)\right].    
\end{equation}
Then, one can make use of the results in Eq.\eqref{Eq:aqmbs_res} to obtain:
\begin{equation}\label{Eq:above}
||\hat\Pi_{W_{n_0}}e^{t\mathcal{L}}(\hat\rho_0)\ket{\psi_{n_0}}\bra{\psi_{n_0}}||^2= \frac{1}{4}\left[1-\frac{4}{L}\cos^2\left(\frac{k}{2}\right) t\sum_j\gamma_j  +O(t^2)\right]\isEquivTo{L\gg 1}\frac{1}{4}\left[1-\frac{8\pi^2}{L^3} t\sum_j\gamma_j + O(t^2)\right] .    
\end{equation}
Since the first-order correction obtained in Eq.~\eqref{Eq:above}, being vanishing in the thermodynamic limit, is manifestly incompatible with an exponentially-fast time decay of the quantifier introduced in Eq.~\eqref{Eq:exp_off_coh} with a finite decay rate in the thermodynamic limit, we deduce the existence of long-lived off-diagonal coherences between an OSQMBS and its orthogonal subspace in the corresponding magnetization sector, and relate them to the existence of OSAQMBS, ie., gapless excitations of the Hamiltonian $\hat H_2$ that governs the decay of off-diagonal coherences in Eq.~\eqref{Eq:coh_off}.

Finally, we remark that it is possible to further simplify Eq.\eqref{Eq:exp_off_coh} by assuming that the exchange terms of the form $\hat S_j^x \hat S_{j+1}^x+\hat S_j^y \hat S_{j+1}^y$ appear only in the dissipative part of the Lindbladian, as done in the main text. In such case, indeed, it is easy to check that $\hat H_1 (\hat H_2)^m \ket{n_0,k} =0,\,\,\forall\, m\geq 0$, since $\hat H_2$ leaves the subspace $\otimes_{j=1}^{L} span\{\ket{-}_j , \ket{+}_j\}$ invariant and such subspace belongs to the kernel of every term of $\hat H_1$ (the latter statement relying crucially on the assumption that the exchange terms do not appear in $\hat H_1$). Therefore, one may write under the above assumption:
\begin{equation}
    ||\hat\Pi_{W_{n_0}}e^{t\mathcal{L}}(\hat\rho_0)\ket{\psi_{n_0}}\bra{\psi_{n_0}}||^2=\frac{1}{4}\bra{n_0,k}e^{-t\hat H_2} \ket{n_0,k},
\end{equation}
as presented in the main text.

\end{document}